\documentclass[times,twocolumn,final]{elsarticle}

 \RequirePackage{geometry}
\usepackage{framed,multirow}
\usepackage{amsmath}
\usepackage{amssymb}
\usepackage{latexsym}
\usepackage{url}
\usepackage{xcolor}
\usepackage{hyperref}
\usepackage{booktabs}
\usepackage{color}
\graphicspath{{./Figs/}}

\def\ie{{\em i.e.}}
\def\etal{{\em et al.}}
\def\eg{{\em e.g.}}

\definecolor{newcolor}{rgb}{.8,.349,.1}

\begin{document}
\begin{frontmatter} 

\title{Preserving Specificity in Federated Graph Learning for fMRI-based Neurological Disorder Identification} 

\author[1]{Junhao {Zhang}}
\author[2]{Qianqian {Wang}}
\author[1]{Xiaochuan {Wang}}
\author[1,3]{Lishan {Qiao}\corref{cor1}}
\author[2]{Mingxia {Liu}\corref{cor1}}

\address[1]{School of Mathematics Science, Liaocheng University, Liaocheng, Shandong, 252000, China. }
\address[2]{Department of Radiology and BRIC, University of North Carolina at Chapel Hill, Chapel Hill, NC, 27599, USA. }
\address[3]{School of Computer Science and Technology, Shandong Jianzhu University, Jinan, Shandong, 250101, China. }

\cortext[cor1]{Corresponding authors: M. Liu (email: mingxia\_liu@med.unc.edu) and L.~Qiao (email: qiaolishan@lcu.edu.cn).}

\begin{abstract}
Resting-state functional magnetic resonance imaging (rs-fMRI) offers a non-invasive approach to examining abnormal brain connectivity associated with brain disorders. 
Graph neural network (GNN) gains popularity in fMRI representation learning and brain disorder analysis with powerful graph representation capabilities. 
Training a general GNN often necessitates a large-scale dataset from multiple imaging centers/sites, but centralizing multi-site data generally faces inherent challenges related to data privacy, security, and storage burden. 
Federated Learning (FL) enables collaborative model training without centralized multi-site fMRI data. 
Unfortunately, previous FL approaches for fMRI analysis often ignore site-specificity, including demographic factors such as age, gender, and education level. 
To this end, we propose a specificity-aware federated graph learning (SFGL) framework for rs-fMRI analysis and automated brain disorder identification, with a server and multiple clients/sites for federated model aggregation and prediction. 
At each client, our model consists of a shared and a personalized branch, where parameters of the shared branch are sent to the server while those of the personalized branch remain local. 
This can facilitate knowledge sharing among sites and also helps preserve site specificity. 
In the shared branch, we employ a spatio-temporal attention graph isomorphism network to learn dynamic fMRI representations. 
In the personalized branch, we integrate vectorized demographic information (\ie, age, gender, and education years) and functional connectivity networks to preserve site-specific characteristics. 
Representations generated by the two branches are then fused for classification. 
Experimental results on two fMRI datasets with a total of 1,218 subjects suggest that SFGL outperforms several state-of-the-art approaches. 
\end{abstract}

\begin{keyword}
Functional MRI, Graph neural network, Federated learning, Demographic information, Specificity
\end{keyword}

\end{frontmatter}

\section{Introduction}\label{S1}

\begin{figure*}[!t]
\setlength{\abovecaptionskip}{0pt}
\setlength{\belowcaptionskip}{0pt}
\setlength\abovedisplayskip{0pt}
\setlength\belowdisplayskip{0pt}
\centering
\includegraphics[width=1\textwidth]{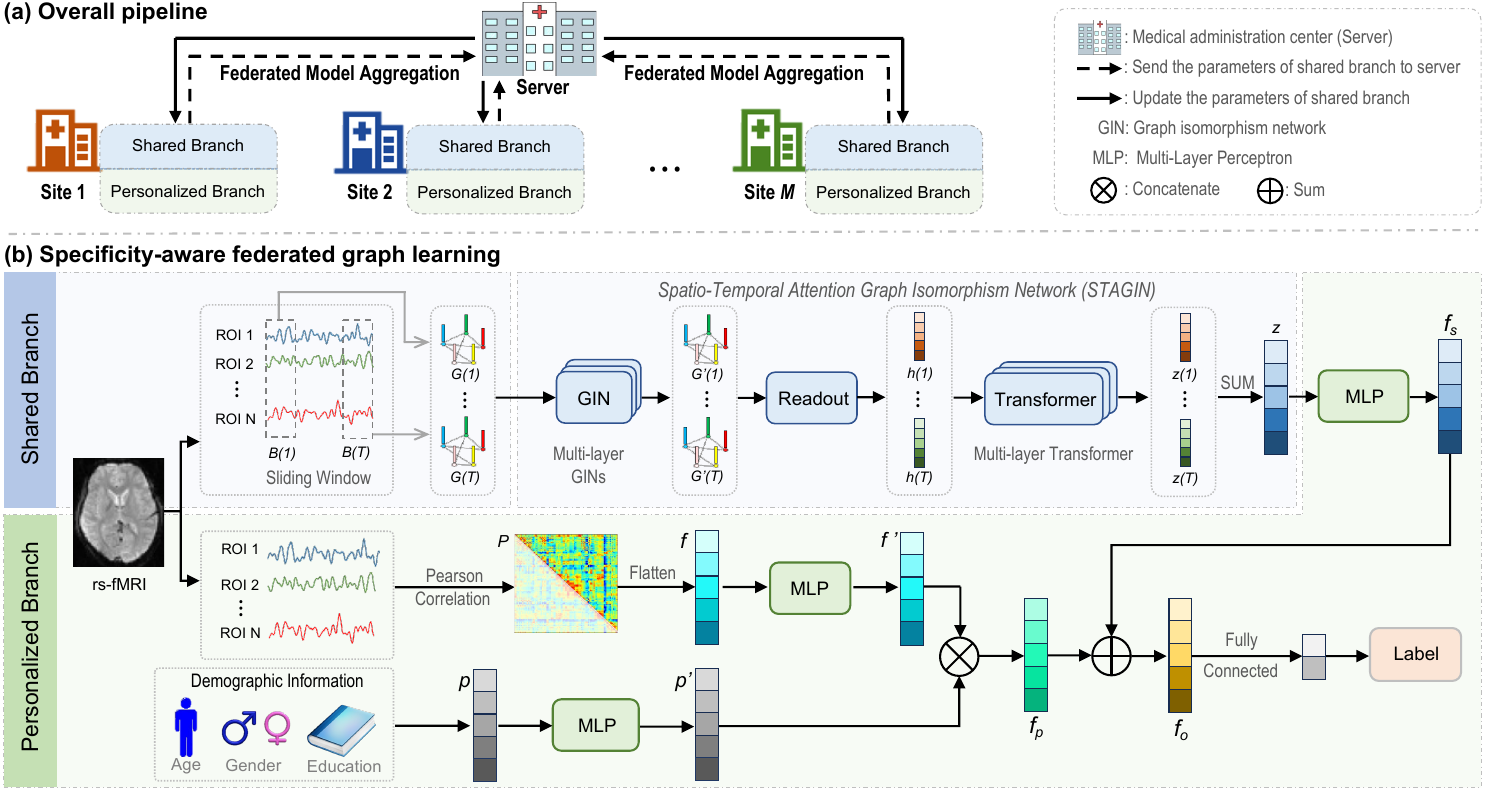}
\caption{Overview of specificity-aware federated graph learning (SFGL) framework, where the top panel is the overall pipeline of our method, involving multiple local clients/sites, a center server, and federated model aggregation, and the bottom panel depicts the specific model architecture at each local client, including a \emph{shared branch} and a \emph{personalized branch}.
In the shared branch, the BOLD signals acquired from rs-fMRI data are segmented using a sliding window to construct dynamic FCNs, followed by a spatio-temporal attention graph isomorphism network (STAGIN) for dynamic graph representation learning. 
In the personalized branch, we first extract vectorized representations of the FCN  and demographic information (\ie, gender, age, and education) respectively, and then concatenate them to obtain the personalized representation for each subject. 
Finally, we aggregate representations learned by shared and personalized branches for brain disorder diagnosis. 
During training, parameters of the shared branch are sent to a server for federated model aggregation, while parameters of the personalized branch remain at each local client to preserve site specificity.}
\label{SFGL}
\end{figure*}

Resting-state functional magnetic resonance imaging (rs-fMRI) serves as a non-invasive tool that can track relevant changes in blood flow, thereby aiding in the identification of abnormal or impaired brain functional connectivity~\cite{khosla2019machine}. 
Many deep learning techniques have been proposed for rs-fMRI analysis and automatic brain disease diagnosis, such as convolutional neural networks (CNNs)~\cite{qureshi20193d,parmar2020deep,lin2022sspnet,xu2023explainable} and graph neural networks (GNNs)~\cite{gadgil2020spatio,gurler2023template,kan2022fbnetgen,qin2022using,chaari2022multigraph,tong2023fmri,jiang2023characterizing}. 
Functional connectivity networks (FCNs) constructed  from rs-fMRI data can be naturally represented as graphs, where each node represents a brain region-of-interest (ROI) and each edge between nodes denotes the connection between two ROIs~\cite{saeidi2022decoding,elgazzar2022benchmarking,jiang2020hi}. 
Given the inherent graph structure of brain FCNs, GNN has shown great potential in fMRI representation learning and brain disorder analysis. 
For example, Tong~\etal~\cite{tong2023fmri} designed a GNN with a low number of parameters for diagnosing Alzheimer's disease and autism. 
But training a GNN model with good generalization ability usually requires a large-scale dataset collected from multiple imaging centers/sites~\cite{zhao2022graph,neyshabur2017exploring,lian2020attention}. 
In addition, conventional GNN methods generally integrate multi-site fMRI data into a central location/server, inevitably bringing challenges related to data privacy, security, and storage burden~\cite{voigt2017eu,act1996health}. 

Recently, federated learning (FL) has been used as an effective decentralized solution to help maintain data privacy and alleviate data storage burdens~\cite{mcmahan2017communication}. 
It allows multiple sites to collaboratively train a shared model without centrally sharing multi-site fMRI data~\cite{pillutla2022federated,li2020federated}. 
For instance, Peng~\etal~\cite{peng2022fedni} proposed an adversarial-based method to predict missing nodes and edges in partial graphs through federated training and constructed a cross-platform GCN node classifier. 
Unfortunately, existing FL approaches for fMRI analysis often ignore the specificity of each imaging site, including demographic factors such as age, gender, and education level of subjects involved in a specific site~\cite{li2020multi}. 

To this end, we propose a specificity-aware federated graph learning (SFGL) framework for fMRI-based neurological disorder identification. 
As illustrated in Fig.~\ref{SFGL}~(a), the proposed SFGL is composed of multiple local clients/sites, a center server, and federated model aggregation. 
At each local client, as shown in Fig.~\ref{SFGL}~(b), the proposed SFGL is composed of a shared branch and a personalized branch, with each client corresponding to a specific imaging site. 
In the shared branch, the fMRI time series is first divided into subsequences across temporal dimensions via a sliding window strategy, followed by dynamic FCN construction for each subject. 
With dynamic FCNs as input, we employ a spatio-temporal attention graph isomorphism network (STAGIN)~\cite{kim2021learning} as the backbone to generate dynamic graph representations. 
In the personalized branch, we first extract vectorized representations of the FCN and demographic information (\ie, gender, age, and education years) respectively, and then concatenate them to obtain the personalized representation for each subject. 
Finally, the dynamic graph representation acquired from the shared branch and the personalized representation obtained from the personalized branch are aggregated, followed by a fully connected layer and a softmax layer for brain disorder identification.  
During model training, the server receives the parameters of the shared branch, whereas the parameters of the personalized branch remain local and are not sent to the central server. 
Therefore, our SFGL not only promotes knowledge exchange among sites without the need for centralized data storage but also maintains site specificity through the personalized branch. 
Experimental results on 1,218 subjects from Autism Brain Imaging Data Exchange (ABIDE)~\cite{di2014autism} and REST-meta-MDD Consortium (REST-MDD)~\cite{yan2019reduced} demonstrate that the SFGL outperforms several state-of-the-art methods in fMRI-based brain disorder diagnosis. 
To the best of our knowledge, this is among the first attempts to incorporate demographic information into federated graph learning for functional MRI analysis and brain disorder identification. 
The major contributions of this work are summarized as follows. 

\begin{itemize}
    \item 
     A novel specificity-aware federated graph learning framework is developed to facilitate cross-site knowledge sharing and preserve site specificity in multi-site fMRI studies without centrally storing data from multiple sites. 
    
    \item 
    A personalized network is designed to employ imaging and non-imaging information in local model training at the client/site side, enabling each site to preserve its own specificity in terms of functional connectivity network and demographic characteristics. 
    
    \item 
    Extensive experiments  on two functional MRI datasets with 1,218 subjects demonstrate that the proposed method outperforms several state-of-the-art methods in fMRI-based brain disorder analysis. 
\end{itemize}

The rest of the paper is organized as follows. 
We briefly review the most relevant work in Section~\ref{S2}. 
In Section~\ref{S3} and Section~\ref{S4}, we respectively describe materials and the proposed method. 
In Section~\ref{S5}, we present experimental setup and results. 
In Section~\ref{S6}, we discuss the influences of several key components and present the limitations of the current work. 
Finally, we conclude the paper in Section~\ref{S7}. 

\section{Related Work}\label{S2}
\subsection{Graph Learning for Functional MRI Analysis}
Resting-state fMRI, as a non-invasive imaging technique, has been widely used for detecting abnormal functional connectivity in the brain. 
Considering that functional connectivity networks (FCNs) derived from fMRI can be naturally represented graphs, some graph learning studies have been proposed for fMRI analysis and brain disorder identification. 
For example, Ktena~\etal~\cite{ktena2018metric} proposed a Siamese GCN framework for learning the similarity between two FCNs with the same topology but different node features. 
Zhang~\etal~\cite{zhang2022classification} proposed a local-to-global GNN model, where the local GNN was used to learn brain network embeddings for each subject, and the global GNN further aggregated the feature representations across subjects. 
Li~\etal~\cite{li2021braingnn} designed an interpretable GNN for fMRI analysis and biomarker detection, where ROI-aware graph convolutional layers were used for node embedding learning and ROI-topK pooling layers could keep high-score ROIs. 
Even though these GNN methods produce promising results in brain disorder identification, they usually ignore temporal dynamics conveyed in fMRI data, which may result in suboptimal performance. 

Recently, some GNN methods have been proposed to capture temporal information in fMRI in addition to spatial patterns. 
For instance, Gadgil~\etal~\cite{gadgil2020spatio} introduced a spatiotemporal graph convolution network that implemented temporal convolutions by constructing natural edges between adjacent time points. 
Cui~\etal~\cite{cui2023personalized} proposed a dual-branch GCN for mild cognitive impairment identification, where spatiotemporal aggregated attention was designed to learn spatial and temporal features of fMRI. 
Behrouz~\etal~\cite{behrouz2022anomaly} treated node embeddings from different GNN layers as hierarchical node states and employed a gated recurrent unit (GRU) to capture temporal properties of functional connectivity networks. 

Given that there are a large number of learnable parameters in GNN models, researchers often utilize multi-site fMRI data to increase sample size for efficient model training. 
However, sending fMRI data acquired from multiple sites to a centralized server is often problematic in clinical practice, which can pose challenges in terms of data privacy and storage burden. 
Therefore, in this work, we propose a federated graph learning framework for multi-site fMRI analysis, which can train a shared model without accessing the local data of each site, thus preserving data privacy and alleviating data storage burden. 

\subsection{Federated Learning for Brain Disease Analsyis}
As a decentralized learning paradigm, federated learning (FL) helps address the challenges of data privacy and storage and storage burden, and has been widely used in many fields such as computer vision and natural language processing. 
In the medical data analysis field, FL also plays a crucial role  because it does not require centralized storage and access to data from individual imaging sites. 
So far, there have been a large number of works utilizing FL for brain disorder analysis. 
For instance, Li~\etal~\cite{li2020multi} proposed a federated domain alignment model for disease diagnosis by sharing local model weights through differential privacy. 
Huang~\etal~\cite{huang2022federated} proposed a federated multi-task learning framework to simultaneously identify multiple related mental disorders. 
They also designed a federated multi-gate mixture of expert classifiers for joint classification of multi-type disorders. 
Fan~\etal~\cite{fan2021federated} utilized guide gradients to update the federated model and trained personalized FL models for diagnostic classification of 3D brain MRI images. 
Zeng~\etal~\cite{zeng2022gradient} introduced a gradient matching federated domain adaptation model for brain disease analysis based on multi-site rs-fMRI data, with a gradient matching loss for pre-training and fine-tuning of both global and local models. 

Unfortunately, existing FL-based methods usually ignore site-specific characteristics of each imaging site in terms of demographic factors (\ie, age, gender, and education) of subjects. 
This hinders the development of site-specific models that are tailored to the unique data characteristics of each site. 
Building upon the advancements and inspirations from these existing methods, we propose a specificity-aware federated graph learning framework, which incorporates the application of demographic information, to generate personalized automatic brain disease diagnosis models based on rs-fMRI data for each site. 

\section{Materials and Data Preprocessing}\label{S3}
\subsection{Materials}
Two public datasets with resting-state functional MRI (rs-fMRI) data are used to validate the effectiveness of our proposed method, including  (1) Autism Brain Imaging Data Exchange (ABIDE)\footnote{\url{http://fcon_1000.projects.nitrc.org/indi/abide}}, and (2) REST-meta-MDD Consortium (REST-MDD)\footnote{\url{http://rfmri.org/REST-meta-MDD}}. 
The demographic information of subjects from these two datasets is provided in Table~\ref{tab_demogra}. 

\begin{table}[!tbp]
\setlength{\belowcaptionskip}{0pt}
\setlength{\abovecaptionskip}{0pt}
\setlength\abovedisplayskip{0pt}
\setlength\belowdisplayskip{0pt}
\renewcommand\arraystretch{1}
\centering
\caption{Demographic and clinical information of subjects from two datasets. The age information is reported as mean$\pm$standard deviation. M: Male; F: Female; ASD: Autism spectrum disorder; MDD: Major depressive disorder; NC: Normal control.}
\scriptsize
\centering
\setlength{\tabcolsep}{0.6pt}
\label{tab_demogra}
\begin{tabular*}{0.48\textwidth}{@{\extracolsep{\fill}} l|cccccc }
\toprule
~~Dataset & Site~ID/Name & Category &Scan~\# & Gender~(M/F) & Age \\
\hline
\multirow{6}{*} {~~ABIDE}
&\multirow{2}{*}{NYU}       &ASD  &$74$	    &$64$/$10$	  &$14.76\pm7.12$   \\
&\multirow{2}{*}            &NC   &$98$     &$72$/$26$	  &$15.75\pm6.19$   \\
\cline{2-6}
&\multirow{2}{*}{UCLA}      &ASD  &$48$     &$42$/$6$     &$13.21\pm2.32$   \\
&\multirow{2}{*}            &NC   &$37$     &$32$/$5$     &$13.03\pm1.97$   \\
\cline{2-6}
&\multirow{2}{*}{UM}        &ASD  &$47$     &$38$/$9$     &$13.71\pm3.37$   \\
&\multirow{2}{*}            &NC   &$73$     &$55$/$18$    &$14.85\pm3.38$   \\
\hline
\multirow{6}{*}{REST-MDD~}
&\multirow{2}{*}{$20$}      &MDD  &$282$	&$99$/$183$	  &$38.74\pm13.74$  \\
&\multirow{2}{*}{}          &NC   &$251$	&$87$/$164$	  &$39.64\pm15.87$  \\
\cline{2-6}
&\multirow{2}{*}{$21$}      &MDD  &$86$     &$38$/$48$    &$34.71\pm12.63$  \\
& \multirow{2}{*}{}         &NC   &$70$     &$31$/$39$    &$36.13\pm12.64$  \\
\cline{2-6}
&\multirow{2}{*}{$25$}      &MDD  &$89$     &$21$/$68$    &$65.60\pm6.75$   \\
&\multirow{2}{*}            &NC   &$63$     &$29$/$34$    &$69.63\pm5.86$   \\
\bottomrule
\end{tabular*}
\end{table}

\subsection{Data Preprocessing}
For ABIDE~\cite{di2014autism}, the top three largest sites are used, including the New York University (NYU), the University of California, Los Angeles (UCLA), and the University of Michigan (UM). 
They respectively include $74$ autism spectrum disorder (ASD) patients and $98$ normal controls (NCs), $48$ ASDs and $37$ NCs, $47$ ASDs and $73$ NCs. 
For the NYU site, fMRI data is collected through a 3T Allegra scanner. 
The parameters are set as follows: repetition time (TR) = $2,000\,ms$, echo time (TE) = $15\,ms$, voxel size = $3.0\times 3.0\times 4.0\,mm^3$, flip angle = $90^\circ $, field-of-view (FOV) = $192\times 240\,mm^2$ and a total of $33$ slices with a thickness of $4\,mm$. 
For the UCLA site, data are obtained through a 3T Trio scanner, with TR = $3,000\,ms$, TE = $28\,ms$, voxel size = $3.0\times 3.0\times 4.0\,mm^3$, flip angle = $90^\circ $, FOV = $192\times 192\,mm^2$ and a total of $34$ slices with a thickness of $4\,mm$. 
For the UM site, data is acquired through a 3T GE Signa scanner, with TR = $2,000\,ms$, TE = $30\,ms$, voxel size = $3.438\times 3.438\times 3.000\,mm^3$, flip angle = $90^\circ $ and a total of $40$ slices with a thickness of $3\,mm$. 
For fMRI data in ABIDE, we preprocess it using the Data Processing Assistant for Resting-State fMRI (DPARSF)~\cite{yan2010dparsf} pipeline. 
Specifically, we first discard the first five volumes to ensure that data collection is in a state of magnetic resonance equilibrium. 
Then, we perform head motion correction, spatial smoothing and normalization, bandpass filtering ($0.01-0.10\,Hz$) of BOLD time series, nuisance signals regression, and spatial standardization of the Montreal Neurological Institute (MNI) template~\cite{tzourio2002automated}. 
Finally, based on the Automated Anatomical Labeling (AAL)~\cite{tzourio2002automated} atlas, each brain is divided into $116$ ROIs, and the average time series are extracted. 

For REST-MDD~\cite{yan2019reduced}, we also select the top three largest sites: Site $20$, Site $21$, and Site $25$. 
They respectively contain $282$ major depressive disorder (MDD) patients  and $251$ NCs, $86$ MDDs and $70$ NCs, and $89$ MDDs and $63$ NCs. 
For Site $20$, fMRI data are acquired through a 3T Trio scanner with a 12-channel receiver coil, with TR = $2,000\,ms$, TE = $30\,ms$, voxel size = $3.44\times 3.44\times 4.00\,mm^3$, flip angle = $90^\circ $, gap = $1.0\,mm$, FOV = $220 \times 220\,mm^2$ and a total of $32$ slices with a thickness of $3\,mm$. 
For Site $21$, data are obtained through a 3T Trio scanner with a 32-channel receiver coil, with TR = $2,000\,ms$, TE = $30\,ms$, voxel size = $3.12\times 3.12\times 4.20\,mm^3$, flip angle = $90^\circ $, gap = $0.7\,mm$, FOV = $200\times 200\,mm^2$ and a total of $33$ slices with a thickness of $3.5\,mm$. 
For Site $25$, data are acquired through a 3T Verio scanner with a 12-channel receiver coil, with TR = $2,000\,ms$, TE = $25\,ms$, voxel size = $3.75\times 3.75\times 4.00\,mm^3$, flip angle = $90^\circ $, gap = $0.0\,mm$, FOV = $240\times 240\,mm^2$ and a total of $36$ slices with a thickness of $4\,mm$. 
For fMRI data in REST-MDD, we also preprocess it using the DPARSF pipeline. 
We first discard the first ten volumes, and then preprocess the data using the same pipeline, including head motion correction, spatial smoothing and normalization, bandpass filtering ($0.01-0.10\,Hz$), nuisance signal regression, and spatial standardization. 
Similarly, all brains are parcellated into $116$ ROIs based on the AAL atlas, leading to average time series of each ROI. 

\section{Methodology}\label{S4}
To collaboratively train a knowledge-sharing model without centralized storage of multi-site data, we propose a specificity-aware federated graph learning (\textbf{SFGL}) framework for fMRI-based neurological disorder identification, which can also preserve the specificity of each client/site. 
As illustrated in Fig.~\ref{SFGL}~(a), the SFGL consists of multiple local clients/sites, a server, and a federated model aggregation strategy. 
For every local client (see Fig.~\ref{SFGL}~(b)), our model consists of a \emph{shared branch} and a \emph{personalized branch}. 
The parameters from the shared branch are transmitted to a server, while the parameters of the personalized branch stay with the local client to maintain site-specific characteristics. 
Details are elaborated as follows. 

\subsection{Shared Branch at Client Side}
As shown in Fig.~\ref{SFGL}~(b), the shared branch primarily consists of two parts: 1) construction of dynamic functional connectivity network/graph sequences via sliding windows and 2) dynamic graph representation learning via a spatio-temporal attention graph isomorphism network. 

\subsubsection{Dynamic Graph Sequence Construction}\label{graph}
The functional connectivity network (FCN) derived from rs-fMRI data can be represented by a graph $G=(\mathcal{V},\mathcal{E})$, where $\mathcal{V}$ represents the set of nodes (ROIs), and $\mathcal{E}$ represents the set of edges (the connection between ROIs). 
The BOLD signal can be defined as {$B=(b_{1},\cdots,b_{N})^{\top}\in R^{N\times D}$}, $b_{i}\in \mathbb{R} ^{D} $, where $N$ is the number of ROIs and  $D$ is the number of time points. 
To capture temporal dynamics within fMRI time series, the BOLD signal can be divided into $T=\left \lfloor \frac{D-\mathit{\Gamma }}{s}+1  \right \rfloor $ segments using the sliding window strategy, with the window size of $\mathit{\Gamma } $ and the stride of $s$. 
Denoting the BOLD signal of the $i$-th ROI at the $t$-th segment  as $b_{i} (t)\in \mathbb{R} ^{\mathit{\Gamma }} (t=1,2,\cdots ,T)$, we utilize Pearson correlation (PC) to construct an FCN at the $t$-th segment. 
The PC coefficient between the BOLD signals of the $i$-th ROI and the $j$-th ROI can be calculated as:
\begin{equation}
p_{ij}(t)=\frac{\mathrm{Cov} (b_i(t),b_j(t))}{\sigma(b_i(t))\sigma(b_j(t))},
\label{EQ1}
\end{equation}
where $\mathrm{Cov}(\cdot , \cdot)$ represents the covariance between the two ROIs, and $\sigma (\cdot )$ is the standard deviation. 
After that, we can obtain a sequence of dynamic brain FCNs for each subject, represented as: $P(t)=(p_{ij}(t)) \in \mathbb{R} ^{N\times N} (t=1,2,\cdots,T)$. 
We evaluate node features by measuring the PC coefficients between each node and its neighbor nodes (that is, each row of $P(t)$), and thus the node feature matrix is represented as $X(t)=P(t)$ at segment $t$. 

The $P(t)$ describes the brain FCN as a dense graph (that is, there exists an edge between each pair of ROIs). 
Previous research has shown that brain connectivity exhibits sparse structure~\cite{sporns2016networks}. 
Therefore, to eliminate noisy/redundant information, we empirically preserve the top $30\%$ of edges with the highest correlations~\cite{kim2021learning} by assigning them a weight value of $1$, while setting the rest to $0$. 
This process allows us to obtain a sparse adjacency matrix $A(t)=\mathbb{I} \left \{ p_{ij}(t)\ge \delta (t) \right \} \in \left \{0,1\right \}$, where $\delta (t)$ represents the selected threshold (\ie, $30\%$). 
Thus, we can represent the constructed graph at the $t$-th time window as $G(t)=(X(t),A(t))$, and  the dynamic graph sequence for each subject can be expressed as $\mathcal{G} =\left \{ G(t) \right \} _{t=1}^T$ ($t=1,\cdots,T$). 

\subsubsection{Dynamic Graph Representation Learning}
As depicted in Fig.~\ref{SFGL}~(b), with the constructed dynamic graph sequence $\mathcal{G}$ as input, we employ a spatio-temporal attention graph isomorphism network (STAGIN)~\cite{kim2021learning} as the backbone for dynamic representation learning, with a multi-layer graph isomorphism network (GIN) and a Transformer. 
GNNs are currently the most popular network framework for analyzing graph-structured data, and they typically consist of two main steps: aggregation and propagation. 
GIN has shown powerful graph representation ability, especially in graph classification tasks~\cite{xu2018powerful}. 
The key difference between GIN and general GNNs lies in the aggregation mechanism and the aggregation mechanism of GIN is formulated as:
\begin{equation}
h_v^{(k)}=\mathrm{MLP} ^{(k)}\left ( \left ( 1+\epsilon ^{(k)} \right ) h_v^{(k-1)}+\sum\nolimits_{u \in \mathcal{N} (v)}^{}h_u^{(k-1)} \right ), 
\label{EQ2}
\end{equation}
where $h_v^{(k)}$ represents the feature vector of node $v$ in the $k$-th layer, $\mathcal{N} (v)$ represents the set of neighbors of node $v$, $\mathrm{MLP} ^{(k)}$ represents a multi-layer perceptron, and $\epsilon ^{(k)}$ is a learnable parameter. 
From Eq.~\ref{EQ2}, it can be observed that GIN can learn adaptive weights for each node, different from the previous average aggregation~\cite{kim2020understanding}. 

In the SFGL, we set the number of GIN layers as $2$. 
For a given time segment, the updated node feature matrix via stacking two GIN layers can be expressed as:
\begin{equation}
X'=\sigma \left ( \epsilon ^{1}I+A \right ) \left ( \sigma \left ( \epsilon ^{0}I+A \right )XW^{0}  \right )W^{1}, 
\label{EQ3}
\end{equation}
where $\sigma$ is the $\mathrm{ReLU} $ activation function, $W^{i}$ is a learnable weight matrix and $\epsilon ^{i}$ is learnable parameter at the $k$-th layer. 

\begin{figure}[!t]
\setlength{\abovecaptionskip}{0pt}
\setlength{\belowcaptionskip}{0pt}
\setlength\abovedisplayskip{0pt}
\setlength\belowdisplayskip{0pt}
\centering
\includegraphics[width=0.49\textwidth]{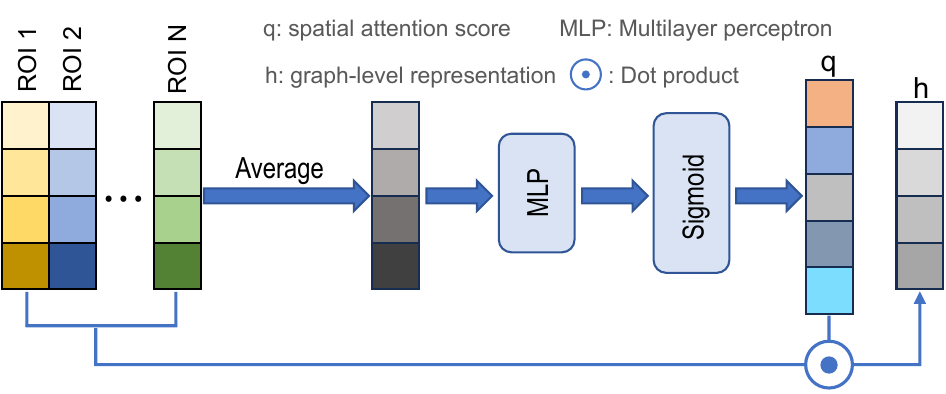}
\caption{Illustration of Sero readout operation, which follows an MLP-based Squeeze-Excitation network.}
\label{Sero}
\end{figure}

To obtain whole-graph-level features, after the GIN layers, the node-level features are aggregated using a readout layer. 
In this work, we employ the Sero readout module proposed by Kim~\etal~\cite{kim2021learning}, which follows a multilayer perceptron (MLP) based squeeze-excitation network.
The Sero module is represented as:
\begin{equation}
q=\mathrm{Sigmoid} (P^{1}\sigma (P^{0}\cdot  \xi (X'))),
\label{EQ4} 
\end{equation}
where $q \in \left ( 0,1 \right )^N $ is the spatial attention vector, $P^{0}$ and $P^{1}$ are learnable matrices, $\sigma$ denotes the $\mathrm{GEReLU} $ activation function, and $\xi(\cdot )$ represents the average operation across the node dimension. 
The details of Sero are illustrated in Fig.~\ref{Sero}. After Sero, we obtain a series of spatial graph representations for each subject, represented as $h(t)=\left \{ X'(t)q(t) \right \}_{t=1}^T$. 

Considering that Transformer has shown remarkable capability in extracting semantic correlations among elements in a long sequence ~\cite{vaswani2017attention}, to capture the graph representations with temporal attention, a single-head Transformer~\cite{vaswani2017attention} with a self-attention mechanism is used to model both short-range and long-range dependencies between different time segments. 
With the generated spatial graph representations $h(t)$ as the input of Transformer, we can obtain a spatiotemporally-attended dynamic graph representation sequence,
represented as $\left \{ z(t) \right \}_{t=1}^T $. 
Finally, we sum the sequence for each subject, yielding a vectorized whole-graph-level representation $z=\sum_{t=1}^{T}z(t)$ for each subject. 

\subsection{Personalized Branch at Client Side}
Previous research has shown that the development of brain disorders is influenced by multiple factors~\cite{liu2017complex,qi2022derivation}. 
Especially, in addition to the imaging information provided by fMRI, demographic information such as gender, age, and education plays a crucial role in understanding the underlying physiological and social relationships~\cite{satizabal2016incidence}. 
In order to preserve the specificity of each site from both imaging and non-imaging perspectives, we design a personalized branch that integrates fMRI features and demographic information simultaneously. 

To extract fMRI features, we first utilize Eq.~\eqref{EQ1} to calculate the PC coefficients between ROIs across the entire time frame, generating the FCN $P$ for each subject. 
Then, we flatten the upper triangular matrix of $P$ into a vector and treat it as the hidden imaging feature $f\in \mathbb{R} ^{N(N-1)/2 }$. 
The final imaging feature for each subject is obtained via an MLP, represented as $f'=fW_f$, where $W_f$ is a learnable matrix. 
For the demographic information (\ie, gender, age, and education years), it is first discretized and digitized to represent the information of each subject as a vector $p$. 
Similarly, the final non-imaging feature for each subject is defined as $p'=pW_p$. 
Finally, we acquire the subject-level representation by concatenating the imaging feature $f'$ and the non-imaging feature $p'$, formulated as:
\begin{equation}
f_p=f'\oplus p',
\label{EQ5}
\end{equation}
where $\oplus$ is the concatenation operation. 

Let $f_s=zW_s$ represent the dynamic representation generated by the shared branch, where $W_s$ is a learnable weight matrix. 
At each client side, we integrate representations generated from the shared and personalized branches to yield a feature vector $f_o$ for each subject, expressed as:
\begin{equation}
f_o=\gamma f_p+(1-\gamma )f_s,
\label{EQ6}
\end{equation}
where $\gamma\in [0,1]$ is used to balance contribution of $f_p$ and $f_s$. 
Subsequently, this feature vector $f_o$ is fed into a fully connected layer and a softmax layer for prediction. 

\subsection{Federated Aggregation at Server Side}
The goal of the SFGL is to facilitate knowledge sharing among different sites via federal learning while preserving site-specific information. 
During the training process, the SFGL only sends parameters of the shared branch to the server side without accessing data of each local client side, thus protecting data privacy and alleviating data storage burden. 
Details of federated aggregation at the server side are introduced as follows. 

For the local model at the $m$-th $(m=1,2,\cdots,M)$ site, the mapping function of the shared branch $\varPhi _{\phi_m}(\cdot )$ is trained using node feature matrix sequence $\left \{ X^m(t) \right \}_{t=1}^T $ and sparse adjacency matrix sequence $\left \{ A^m(t) \right \}_{t=1}^T $ of $m$-th site as input. 
Here, $\phi _m$ represents the learnable parameters of $\varPhi (\cdot )$, and the output can be represented as:
\begin{equation}
z^m=\varPhi _{\phi_m}\left ( \left \{ X^m(t) \right \}_{t=1}^T ,\left \{ A^m(t) \right \}_{t=1}^T \right ).
\label{EQ7}
\end{equation}
Additionally, the mapping function of the personalized branch $\Theta _{\theta _m}(\cdot )$ is trained using both imaging features $f^m$ and vectorized demographic information representations $p^m$ of $m$-th site as input. 
Here, $\theta _{m}$ represents the learnable parameters of $\Theta (\cdot )$, and the output is expressed as:
\begin{equation}
f_{o}^m=\Theta _{\theta _m}\left ( f^m,p^m,z^m \right ).
\label{EQ8}
\end{equation}

The vector $f_{o}$ is further fed into a fully connected layer and a Softmax layer for prediction in the personalized branch. 
Therefore, the cross-entropy loss of each local model is defined as: 
\begin{equation}
L_c^m=-\sum\nolimits_{i\in Y^m}^{} \left ( y_i^m\log_{}{\left ( g_i^m \right ) }+\left ( 1-y_i^m \right )\log_{}{\left ( 1-g_i^m \right ) } \right ),
\label{EQ9}
\end{equation}
where $Y^m$ represents the set of data at the $m$-th client, $y_i^m$ is the label of the $i$-th sample and $g_i^m$ is the predicted probability of the $i$-th sample at the $m$-th client. 

Due to the norm-preserving property of orthogonal matrices, we aim to encourage orthogonality of feature matrices in the SFGL to prevent gradient explosion or vanishing during the optimization process~\cite{brock2016neural}. 
Therefore, we introduce an orthogonal constraint loss for the feature matrices as follows:
\begin{equation}
L_{ortho}^m=\left \| \frac{1}{\mu _m}{X^{'m}}^{\mathrm{T}}X^{'m}-I \right \|_2,
\label{EQ10}
\end{equation}
where $\mu _m=\mathrm{max}\left ( {X^{'m}}^{\mathrm{T}}{X^{'m}} \right )$. 
Therefore, the final loss for the $m$-th client is defined as
\begin{equation}
L^m=L_c^m+\lambda L_{ortho}^m ,
\label{EQ11}
\end{equation}
where $\lambda$ is a hyperparameter that balances $L_c^m$ and $L_{ortho}^m$. 

During the $r$-th round of client-server communication, the following parameter update procedure is executed: 
\begin{equation}
\phi _m^{r+1} \gets \phi  _m^{r}-\eta \nabla L^m\left ( \left \{ X^m(t) \right \}_{t=1}^T,\left \{ A^m(t) \right \}_{t=1}^T,f^m,p^m,\mathcal{Y}^m \right ) ,
\label{EQ12}
\end{equation}
\begin{equation}
\theta _m^{r+1} \gets \theta  _m^{r}-\eta \nabla L^m\left ( \left \{ X^m(t) \right \}_{t=1}^T,\left \{ A^m(t) \right \}_{t=1}^T,f^m,p^m,\mathcal{Y}^m \right ) ,
\label{EQ13}
\end{equation}
where $\eta $ is learning rate, and $\mathcal{Y} ^m$ represents the set of category labels of subjects for the $m$-th client. 

At each client side, the server receives the parameters $\phi _{m}^{r+1}$ from the shared branch, while the parameters $\theta_{m}^{r+1}$ of the personalized branch are kept local and only updated on-site. 
On the server side, we conduct weighted aggregation of $\left \{ \phi _m^{r+1} \right \}_{m=1}^M $ from all $M$ sites as: 
\begin{equation}
\phi^{r+1}=\sum\nolimits_{m=1}^{M}\frac{n_m}{n}\phi_m^{r+1},
\label{EQ14}
\end{equation}
where $n_m$ is the number of samples at the $m$-th client and $n=\sum n_m$. 
After receiving the aggregated parameter $\phi ^{r+1}$, each client updates its own parameter $\phi _m^{r+1}$ and commences the next round of communication. 

\subsection{Implementation Details}
The proposed SFGL is implemented using PyTorch 3.8 on NVIDIA GeForce RTX 2060 SUPER and NVIDIA GeForce RTX 3070 Ti. 
We set the communication round $R=10$ for server-client communication, local training epoch $E=5$ at each client, sliding window length $\mathit{\Gamma}=30$, sliding window stride $s=2$, balancing coefficient $\gamma =0.8$ in Eq.~\eqref{EQ6} and regularization factor $\lambda =1\times 10^{-5}$ in Eq.~\eqref{EQ11}. 
In addition, we set the batch size to $4$ and the dropout rate to $0.5$. 
This model is optimized using the Adam with a learning rate of $\eta=0.001$.  
In Section~\ref{S6}, we will investigate the influence of several important hyperparameters on the experimental results. 

\begin{table*}[!tbp] 
\setlength{\belowcaptionskip}{0pt}
\setlength{\abovecaptionskip}{0pt}
\setlength\abovedisplayskip{0pt}
\setlength\belowdisplayskip{0pt}
\scriptsize
\tiny
\centering
\renewcommand{\arraystretch}{1.1}
\setlength{\tabcolsep}{2pt}	 
\caption{Results (\%) of eleven different methods in ASD vs. NC classification on the ABIDE dataset in terms of mean$\pm$standard deviation. ``tr\_\textless site\textgreater'' represents the site used for training in the Cross operation. Best results are shown in bold.} 
\begin{tabular*}{1\textwidth}{@{\extracolsep{\fill}} l|ccccc|ccccc|ccccc|ccccc}
\toprule
\multirow{2}{*}{~Method}
& \multicolumn{5}{c|}{NYU} & \multicolumn{5}{c|}{UCLA} & \multicolumn{5}{c|}{UM} 
& \multicolumn{5}{c}{Average} \\                                                   
\cline{2-21} 
&ACC & PRE & REC & AUC & F1 & ACC & PRE & REC & AUC & F1 & ACC & PRE & REC & AUC & F1 & ACC & PRE & REC & AUC & F1 \\ 
\hline
\multirow{1}{*}{~tr\_NYU}
& -  & - & - & - & - 
& \shortstack[c]{58.8$\pm$5.1} & \shortstack[c]{51.8$\pm$6.9} & \textbf{\shortstack[c]{75.7$\pm$23.8}} 
& \shortstack[c]{68.1$\pm$7.4} & \textbf{\shortstack[c]{61.5$\pm$10.4}} & \shortstack[c]{62.5$\pm$3.1} 
& \shortstack[c]{67.5$\pm$6.0} & \shortstack[c]{73.9$\pm$20.5} & \shortstack[c]{58.6$\pm$5.4} 
& \shortstack[c]{70.6$\pm$8.1} & \shortstack[c]{60.6$\pm$4.1} & \shortstack[c]{59.6$\pm$6.5}
& \textbf{\shortstack[c]{74.8$\pm$22.2}} & \shortstack[c]{63.3$\pm$6.4} & \shortstack[c]{66.0$\pm$9.3}\\

\multirow{1}{*}{~tr\_UCLA}
& \shortstack[c]{54.0$\pm$2.8} & \shortstack[c]{65.6$\pm$7.6} & \shortstack[c]{42.9$\pm$23.8} 
& \shortstack[c]{59.5$\pm$2.1} & \shortstack[c]{51.9$\pm$12.7} 
& -  & - & - & - & - 
& \shortstack[c]{60.0$\pm$4.9} 
& \shortstack[c]{69.8$\pm$3.3} & \shortstack[c]{60.2$\pm$17.6} & \shortstack[c]{60.4$\pm$4.0} 
& \shortstack[c]{64.7$\pm$9.0} & \shortstack[c]{57.0$\pm$3.8} & \shortstack[c]{67.7$\pm$5.5} 
& \shortstack[c]{51.6$\pm$20.7} & \shortstack[c]{60.0$\pm$3.1} & \shortstack[c]{58.3$\pm$10.8}\\

\multirow{1}{*}{~tr\_UM}
& \shortstack[c]{57.7$\pm$3.9} & \shortstack[c]{61.5$\pm$2.9} & \shortstack[c]{68.4$\pm$12.5} 
& \shortstack[c]{56.9$\pm$5.7} & \shortstack[c]{64.7$\pm$4.5} & \shortstack[c]{57.1$\pm$4.1} 
& \shortstack[c]{50.9$\pm$2.9} & \shortstack[c]{75.6$\pm$7.4} & \shortstack[c]{60.5$\pm$2.9} 
& \shortstack[c]{60.8$\pm$1.1} 
& -  & - & - & - & - 
& \shortstack[c]{57.4$\pm$4.0} & \shortstack[c]{56.2$\pm$2.9} 
& \shortstack[c]{66.2$\pm$10.0} & \shortstack[c]{58.7$\pm$4.3} & \shortstack[c]{62.8$\pm$2.8}\\

\multirow{1}{*}{~Single}
& \shortstack[c]{60.5$\pm$12.1} & \shortstack[c]{67.0$\pm$15.7} & \shortstack[c]{60.2$\pm$7.3} 
& \shortstack[c]{61.1$\pm$14.0} & \shortstack[c]{63.4$\pm$8.9} & \shortstack[c]{60.0$\pm$13.6} 
& \shortstack[c]{53.8$\pm$25.6} & \shortstack[c]{56.8$\pm$29.7} & \shortstack[c]{57.4$\pm$18.1} 
& \shortstack[c]{55.3$\pm$27.2} & \shortstack[c]{61.7$\pm$12.5} & \shortstack[c]{67.5$\pm$11.8} 
& \shortstack[c]{71.2$\pm$14.6} & \shortstack[c]{60.3$\pm$7.0} & \shortstack[c]{69.3$\pm$10.7}
& \shortstack[c]{60.7$\pm$12.7} & \shortstack[c]{62.8$\pm$17.7} & \shortstack[c]{62.7$\pm$17.2}
& \shortstack[c]{59.6$\pm$13.0} & \shortstack[c]{62.7$\pm$15.6}\\

\multirow{1}{*}{~Mix}
& -&-&-&-&-
& -&-&-&-&-
& -&-&-&-&-
& \shortstack[c]{60.7$\pm$2.4} & \shortstack[c]{61.7$\pm$3.2} & \shortstack[c]{63.0$\pm$12.6}   
& \shortstack[c]{60.5$\pm$2.8} & \shortstack[c]{68.1$\pm$4.2}\\

\hline
\multirow{1}{*}{~FedAvg}
& \shortstack[c]{67.5$\pm$2.9} & \shortstack[c]{69.8$\pm$2.9} & \shortstack[c]{75.5$\pm$8.5} 
& \shortstack[c]{71.4$\pm$7.6} & \shortstack[c]{72.5$\pm$3.9} & \shortstack[c]{61.2$\pm$4.4} 
& \shortstack[c]{57.7$\pm$6.3} & \shortstack[c]{40.5$\pm$14.6} & \shortstack[c]{61.7$\pm$9.3} 
& \shortstack[c]{47.6$\pm$13.0} & \shortstack[c]{65.8$\pm$7.9} & \shortstack[c]{75.8$\pm$5.0} 
& \shortstack[c]{64.3$\pm$20.2} & \shortstack[c]{68.4$\pm$4.9} & \shortstack[c]{69.6$\pm$13.9}
& \shortstack[c]{64.8$\pm$5.1} & \shortstack[c]{70.1$\pm$4.8} & \shortstack[c]{65.3$\pm$14.4}
& \shortstack[c]{69.0$\pm$7.2} & \shortstack[c]{67.7$\pm$10.2}\\
 
\multirow{1}{*}{~FedProx}
& \shortstack[c]{67.5$\pm$11.5} & \shortstack[c]{68.4$\pm$11.3} & \shortstack[c]{76.0$\pm$17.5} 
& \shortstack[c]{69.2$\pm$10.0} & \shortstack[c]{73.6$\pm$9.9} & \shortstack[c]{58.8$\pm$5.3} 
& \shortstack[c]{55.0$\pm$24.5} & \shortstack[c]{23.0$\pm$19.8} & \shortstack[c]{61.3$\pm$8.6} 
& \shortstack[c]{38.6$\pm$18.4} & \textbf{\shortstack[c]{70.0$\pm$9.4}} & \textbf{\shortstack[c]{75.9$\pm$13.0}} 
& \shortstack[c]{60.3$\pm$14.3} & \shortstack[c]{69.6$\pm$8.0} & \shortstack[c]{67.2$\pm$9.4}
& \shortstack[c]{65.4$\pm$8.7} & \shortstack[c]{66.4$\pm$16.3} & \shortstack[c]{53.1$\pm$17.2}
& \shortstack[c]{66.7$\pm$8.9} & \shortstack[c]{59.8$\pm$12.6}\\

\multirow{1}{*}{~MOON}
& \shortstack[c]{65.7$\pm$5.7} & \shortstack[c]{62.4$\pm$7.9} & \shortstack[c]{69.4$\pm$16.5} 
& \shortstack[c]{59.8$\pm$4.5} & \shortstack[c]{65.7$\pm$6.4} & \shortstack[c]{62.3$\pm$10.9} 
& \shortstack[c]{42.3$\pm$7.4} & \shortstack[c]{64.8$\pm$19.5} & \shortstack[c]{56.4$\pm$10.7} 
& \shortstack[c]{53.3$\pm$11.2} & \shortstack[c]{65.8$\pm$4.9} & \shortstack[c]{73.8$\pm$7.9} 
& \shortstack[c]{65.7$\pm$21.0} & \shortstack[c]{68.3$\pm$5.3} & \shortstack[c]{70.0$\pm$16.3}
& \shortstack[c]{64.1$\pm$7.1} & \shortstack[c]{61.7$\pm$7.8} & \shortstack[c]{67.3$\pm$19.0}
& \shortstack[c]{62.2$\pm$6.8} & \shortstack[c]{64.4$\pm$11.3}\\

\multirow{1}{*}{~pFedMe}
& \shortstack[c]{66.8$\pm$7.3} & \shortstack[c]{70.8$\pm$9.2} & \shortstack[c]{70.0$\pm$24.3} 
& \shortstack[c]{66.7$\pm$8.5} & \shortstack[c]{70.1$\pm$14.0} & \shortstack[c]{61.7$\pm$8.2} 
& \shortstack[c]{56.7$\pm$10.5} & \shortstack[c]{46.0$\pm$14.5} & \textbf{\shortstack[c]{68.3$\pm$12.0}} 
& \shortstack[c]{50.7$\pm$8.8} & \shortstack[c]{65.0$\pm$6.9} & \shortstack[c]{69.0$\pm$8.5} 
& \shortstack[c]{73.4$\pm$9.6} & \shortstack[c]{63.5$\pm$9.3} & \shortstack[c]{73.8$\pm$6.3}
& \shortstack[c]{64.5$\pm$7.5} & \shortstack[c]{68.1$\pm$9.4} & \shortstack[c]{68.7$\pm$16.1}
& \shortstack[c]{66.8$\pm$9.9} & \shortstack[c]{68.4$\pm$9.7}\\

\multirow{1}{*}{~LGFed}
& \shortstack[c]{67.0$\pm$11.3} & \shortstack[c]{67.5$\pm$8.6} & \shortstack[c]{78.5$\pm$15.0} 
& \shortstack[c]{69.4$\pm$14.2} & \shortstack[c]{72.6$\pm$10.4} & \shortstack[c]{58.8$\pm$12.9} 
& \shortstack[c]{52.8$\pm$21.9} & \shortstack[c]{51.3$\pm$25.1} & \shortstack[c]{57.1$\pm$17.0} 
& \shortstack[c]{52.0$\pm$18.2} & \shortstack[c]{66.0$\pm$4.9} & \shortstack[c]{69.5$\pm$8.1} 
& \shortstack[c]{71.0$\pm$10.5} & \shortstack[c]{65.6$\pm$4.7} & \shortstack[c]{73.5$\pm$3.3}
& \shortstack[c]{63.9$\pm$10.0} & \shortstack[c]{65.9$\pm$12.8} & \shortstack[c]{67.0$\pm$16.8}
& \shortstack[c]{66.5$\pm$12.0} & \shortstack[c]{69.5$\pm$10.7}\\

\hline
\multirow{1}{*}{~SFGL (Ours)}
& \textbf{\shortstack[c]{70.4$\pm$8.9}} & \textbf{\shortstack[c]{72.0$\pm$4.7}} & \textbf{\shortstack[c]{78.6$\pm$15.9}} 
& \textbf{\shortstack[c]{74.3$\pm$7.1}} & \textbf{\shortstack[c]{75.1$\pm$10.2}} & \textbf{\shortstack[c]{64.0$\pm$6.8}} 
& \textbf{\shortstack[c]{60.0$\pm$20.3}} & \shortstack[c]{48.6$\pm$20.7} & \shortstack[c]{60.8$\pm$11.8} 
& \shortstack[c]{53.7$\pm$13.7} & \textbf{\shortstack[c]{70.0$\pm$5.5}} & \shortstack[c]{75.3$\pm$7.4} 
& \textbf{\shortstack[c]{75.3$\pm$15.7}} & \textbf{\shortstack[c]{71.8$\pm$6.7}} & \textbf{\shortstack[c]{75.3$\pm$7.0}}
& \textbf{\shortstack[c]{68.1$\pm$7.1}} & \textbf{\shortstack[c]{71.4$\pm$11.7}} & \shortstack[c]{72.1$\pm$17.4}
& \textbf{\shortstack[c]{71.9$\pm$8.5}} & \textbf{\shortstack[c]{71.8$\pm$10.3}}\\

\bottomrule
\end{tabular*}
\label{ABIDE_result}
\end{table*}

\begin{table*}[!tbp] 
\setlength{\belowcaptionskip}{0pt}
\setlength{\abovecaptionskip}{0pt}
\setlength\abovedisplayskip{0pt}
\setlength\belowdisplayskip{0pt}
\scriptsize
\tiny
\centering
\renewcommand{\arraystretch}{1.1}
\setlength{\tabcolsep}{2pt}	 
\caption{Results (\%) of eleven different methods in MDD vs. NC classification on the REST-MDD dataset in terms of mean$\pm$standard deviation. ``tr\_\textless site\textgreater'' represents the site used for training in the Cross operation. Best results are shown in bold.} 
\begin{tabular*}{1\textwidth}{@{\extracolsep{\fill}} l|ccccc|ccccc|ccccc|ccccc}
\toprule
\multirow{2}{*}{~Method}
& \multicolumn{5}{c|}{Site20} & \multicolumn{5}{c|}{Site21} & \multicolumn{5}{c|}{Site25} 
& \multicolumn{5}{c}{Average} \\
                                                   
\cline{2-21} 
&ACC & PRE & REC & AUC & F1 & ACC & PRE & REC & AUC & F1 & ACC & PRE & REC & AUC & F1 & ACC & PRE & REC & AUC & F1 \\ 
\hline
\multirow{1}{*}{~tr\_Site20}
& -  & - & - & - & - 
& \shortstack[c]{59.6$\pm$3.4} & \shortstack[c]{60.7$\pm$5.7} & \shortstack[c]{60.5$\pm$14.6}
& \shortstack[c]{56.6$\pm$4.0} & \shortstack[c]{57.4$\pm$7.8} & \shortstack[c]{56.0$\pm$5.2} 
& \shortstack[c]{52.8$\pm$6.0} & \shortstack[c]{65.2$\pm$7.0} & \shortstack[c]{55.7$\pm$4.2} 
& \shortstack[c]{58.4$\pm$6.6} & \shortstack[c]{57.8$\pm$4.2} & \shortstack[c]{56.7$\pm$5.8}
& \shortstack[c]{62.8$\pm$10.8} & \shortstack[c]{56.2$\pm$4.1} & \shortstack[c]{57.9$\pm$7.2}\\

\multirow{1}{*}{~tr\_Site21}
& \shortstack[c]{54.7$\pm$2.0} & \shortstack[c]{53.7$\pm$2.1} & \shortstack[c]{66.1$\pm$25.9} 
& \shortstack[c]{54.1$\pm$1.5} & \shortstack[c]{62.2$\pm$14.2} 
& -  & - & - & - & - 
& \shortstack[c]{59.2$\pm$1.1} 
& \shortstack[c]{60.3$\pm$7.2} & \shortstack[c]{68.7$\pm$19.0} & \shortstack[c]{60.6$\pm$2.7}
& \shortstack[c]{61.8$\pm$7.3} & \shortstack[c]{56.9$\pm$1.6} & \shortstack[c]{57.0$\pm$4.6}
& \shortstack[c]{67.4$\pm$22.5} & \shortstack[c]{57.3$\pm$2.1} & \shortstack[c]{62.0$\pm$10.7}\\

\multirow{1}{*}{~tr\_Site25}
& \shortstack[c]{54.8$\pm$1.1} & \shortstack[c]{57.5$\pm$0.9} & \shortstack[c]{56.0$\pm$6.2}
& \shortstack[c]{54.0$\pm$0.9} & \shortstack[c]{56.7$\pm$2.9} & \shortstack[c]{58.3$\pm$3.5}
& \shortstack[c]{63.3$\pm$3.6} & \shortstack[c]{58.1$\pm$7.3} & \shortstack[c]{57.3$\pm$2.7}
& \shortstack[c]{60.6$\pm$2.7} 
& -  & - & - & - & - 
& \shortstack[c]{56.6$\pm$2.3} & \shortstack[c]{60.4$\pm$2.2} 
& \shortstack[c]{57.1$\pm$6.7} & \shortstack[c]{55.6$\pm$1.8} & \shortstack[c]{58.6$\pm$2.8}\\

\multirow{1}{*}{~Single}
& \shortstack[c]{55.6$\pm$3.9} & \shortstack[c]{57.4$\pm$3.2} & \shortstack[c]{62.1$\pm$16.0} 
& \shortstack[c]{56.7$\pm$4.5} & \shortstack[c]{59.6$\pm$8.3} & \shortstack[c]{59.0$\pm$5.0} 
& \shortstack[c]{61.4$\pm$3.5} & \textbf{\shortstack[c]{68.6$\pm$12.8}} & \shortstack[c]{57.3$\pm$8.3} 
& \textbf{\shortstack[c]{64.8$\pm$6.6}} & \shortstack[c]{58.0$\pm$10.7} & \shortstack[c]{62.7$\pm$9.0} 
& \shortstack[c]{66.3$\pm$9.7} & \shortstack[c]{59.4$\pm$10.0} & \shortstack[c]{64.4$\pm$8.6}
& \shortstack[c]{57.5$\pm$6.5} & \shortstack[c]{60.5$\pm$5.2} & \shortstack[c]{65.6$\pm$12.8}
& \shortstack[c]{57.8$\pm$7.6} & \shortstack[c]{62.9$\pm$7.8}\\

\multirow{1}{*}{~Mix}
& -&-&-&-&-
& -&-&-&-&-
& -&-&-&-&-
& \shortstack[c]{58.0$\pm$3.9} & \shortstack[c]{61.5$\pm$0.8} & \shortstack[c]{60.6$\pm$17.5}
& \shortstack[c]{60.6$\pm$3.0} & \shortstack[c]{61.1$\pm$11.5}\\

\hline
\multirow{1}{*}{~FedAvg}
& \shortstack[c]{60.0$\pm$3.9} & \shortstack[c]{60.4$\pm$3.7} & \shortstack[c]{68.8$\pm$6.6} 
& \shortstack[c]{63.8$\pm$3.2} & \shortstack[c]{64.3$\pm$3.5} & \shortstack[c]{59.0$\pm$6.4} 
& \shortstack[c]{63.7$\pm$4.4} & \shortstack[c]{59.3$\pm$15.6} & \shortstack[c]{61.5$\pm$8.5} 
& \shortstack[c]{61.4$\pm$9.3} & \shortstack[c]{61.2$\pm$4.7} & \shortstack[c]{69.2$\pm$4.7} 
& \shortstack[c]{60.7$\pm$11.7} & \shortstack[c]{62.5$\pm$7.8} & \shortstack[c]{64.7$\pm$6.2}
& \shortstack[c]{60.0$\pm$5.0} & \shortstack[c]{62.4$\pm$4.3} & \shortstack[c]{65.4$\pm$11.3}
& \shortstack[c]{62.8$\pm$6.5} & \shortstack[c]{63.9$\pm$6.4}\\
 
\multirow{1}{*}{~FedProx}
& \shortstack[c]{63.6$\pm$3.4} & \shortstack[c]{63.9$\pm$5.0} & \shortstack[c]{71.6$\pm$13.7} 
& \shortstack[c]{67.1$\pm$4.2} & \shortstack[c]{67.5$\pm$5.3} & \shortstack[c]{60.0$\pm$8.0} 
& \shortstack[c]{65.7$\pm$6.6} & \shortstack[c]{58.1$\pm$13.6} & \shortstack[c]{64.0$\pm$6.9} 
& \shortstack[c]{61.7$\pm$10.8} & \shortstack[c]{59.2$\pm$4.1} & \shortstack[c]{68.0$\pm$8.2} 
& \shortstack[c]{57.3$\pm$16.1} & \shortstack[c]{63.7$\pm$5.6} & \shortstack[c]{62.2$\pm$9.1}
& \shortstack[c]{61.0$\pm$5.1} & \shortstack[c]{64.8$\pm$6.6} & \shortstack[c]{66.3$\pm$14.5}
& \shortstack[c]{65.7$\pm$5.6} & \shortstack[c]{65.6$\pm$8.4}\\

\multirow{1}{*}{~MOON}
& \shortstack[c]{61.7$\pm$2.6} & \shortstack[c]{63.2$\pm$3.6} & \shortstack[c]{65.9$\pm$11.9} 
& \shortstack[c]{67.5$\pm$2.7} & \shortstack[c]{64.6$\pm$5.0} & \shortstack[c]{61.5$\pm$10.0} 
& \shortstack[c]{69.0$\pm$8.1} & \shortstack[c]{56.9$\pm$14.8} & \textbf{\shortstack[c]{68.9$\pm$10.6}} 
& \shortstack[c]{62.4$\pm$11.7} & \shortstack[c]{59.3$\pm$9.8} & \textbf{\shortstack[c]{72.1$\pm$4.3}} 
& \shortstack[c]{49.4$\pm$23.4} & \shortstack[c]{62.0$\pm$11.0} & \shortstack[c]{58.7$\pm$20.1}
& \shortstack[c]{60.8$\pm$7.4} & \textbf{\shortstack[c]{65.5$\pm$5.3}} & \shortstack[c]{61.1$\pm$16.7}
& \shortstack[c]{66.2$\pm$8.1} & \shortstack[c]{63.2$\pm$12.2}\\

\multirow{1}{*}{~pFedMe}
& \shortstack[c]{61.4$\pm$0.1} & \shortstack[c]{58.7$\pm$1.8} & \shortstack[c]{62.4$\pm$13.9} 
& \shortstack[c]{59.4$\pm$3.6} & \shortstack[c]{60.5$\pm$5.4} & \shortstack[c]{60.0$\pm$8.4} 
& \textbf{\shortstack[c]{67.1$\pm$11.2}} & \shortstack[c]{52.3$\pm$21.2} & \shortstack[c]{62.5$\pm$8.7} 
& \shortstack[c]{58.8$\pm$15.7} & \shortstack[c]{60.6$\pm$9.3} & \shortstack[c]{69.2$\pm$12.3} 
& \shortstack[c]{50.6$\pm$17.0} & \shortstack[c]{63.1$\pm$9.5} & \shortstack[c]{58.4$\pm$13.3}
& \shortstack[c]{60.7$\pm$6.2} & \shortstack[c]{61.5$\pm$8.4} & \shortstack[c]{58.2$\pm$17.3}
& \shortstack[c]{60.2$\pm$7.3} & \shortstack[c]{59.8$\pm$11.5}\\

\multirow{1}{*}{~LGFed}
& \shortstack[c]{60.4$\pm$4.9} & \shortstack[c]{62.7$\pm$6.0} & \shortstack[c]{62.0$\pm$12.1} 
& \shortstack[c]{65.5$\pm$4.4} & \shortstack[c]{62.4$\pm$6.7} & \shortstack[c]{56.4$\pm$6.9} 
& \shortstack[c]{61.0$\pm$5.8} & \shortstack[c]{58.1$\pm$10.9} & \shortstack[c]{58.5$\pm$8.0} 
& \shortstack[c]{59.5$\pm$5.1} & \shortstack[c]{61.8$\pm$9.9} & \shortstack[c]{66.7$\pm$5.2} 
& \shortstack[c]{69.7$\pm$10.5} & \shortstack[c]{63.3$\pm$6.8} & \shortstack[c]{68.1$\pm$6.2}
& \shortstack[c]{59.5$\pm$7.3} & \shortstack[c]{63.1$\pm$5.7} & \shortstack[c]{62.8$\pm$11.2}
& \shortstack[c]{63.9$\pm$6.4} & \shortstack[c]{63.3$\pm$6.0}\\

\hline
\multirow{1}{*}{~SFGL (Ours)}
& \textbf{\shortstack[c]{64.4$\pm$2.4}} & \textbf{\shortstack[c]{64.5$\pm$2.9}} & \textbf{\shortstack[c]{72.7$\pm$10.2}} 
& \textbf{\shortstack[c]{68.9$\pm$4.1}} & \textbf{\shortstack[c]{68.3$\pm$4.1}} & \textbf{\shortstack[c]{62.0$\pm$4.3}} 
& \shortstack[c]{63.7$\pm$4.2} & \shortstack[c]{65.1$\pm$7.6} & \shortstack[c]{62.3$\pm$6.9} 
& \shortstack[c]{64.4$\pm$4.1} & \textbf{\shortstack[c]{62.5$\pm$6.0}} & \shortstack[c]{64.5$\pm$3.4} 
& \textbf{\shortstack[c]{79.8$\pm$11.4}} & \textbf{\shortstack[c]{69.2$\pm$7.9}} & \textbf{\shortstack[c]{71.4$\pm$5.6}}
& \textbf{\shortstack[c]{62.9$\pm$4.7}} & \shortstack[c]{64.3$\pm$3.4} & \textbf{\shortstack[c]{72.6$\pm$9.7}}
& \textbf{\shortstack[c]{67.8$\pm$6.3}} & \textbf{\shortstack[c]{68.2$\pm$4.6}}\\

\bottomrule
\end{tabular*}
\label{MDD_result}
\end{table*}

\section{Experiments}\label{S5}
\subsection{Experimental Settings}
A 5-fold cross-validation (CV) strategy is used in the experiments. 
Specifically, the data at each site is randomly divided into five mutually exclusive subsets. 
Four subsets are used as training set, while the remaining subset is used as test set in turn. 
For each site, we record the average and standard deviation of the classification results from the five folds. 
Five evaluation metrics are used: accuracy (ACC), precision (PRE), recall (REC), area under the ROC curve (AUC) and F1-Score (F1). 

\subsection{Methods for Comparison}
The proposed SFGL is compared with two types of approaches, including (1) three non-FL strategies (\ie, Cross, Single, and Mix) that directly merge the shared branch and the personalized branch of our SFGL into a unified workflow, and (2) five state-of-the-art FL methods (\ie, FedAvg, FedProx, MOON, pFedMe, and LGFed) that use specific federated learning strategies and the similar network architecture as SFGL. 
Details of these competing methods are introduced as follows. 

(1) \textbf{Cross}. This method uses data from one site as the training set and the data from the remaining sites serves as an independent test set. 
We use ``tr\_\textless site\textgreater'' to represent the method, where the term ``site'' means the imaging site used for model training.

(2) \textbf{Single}. This method performs training and test separately using data from a single site through the 5-fold CV strategy. 
That is, there is no knowledge transfer/sharing among these individual sites/clients. 

(3) \textbf{Mix}. In this method, data from all sites are mixed for training and test through the 5-fold CV strategy. 

(4) \textbf{FedAvg}~\cite{mcmahan2017communication}. For a fair comparison, this method has the same network architecture as our SFGL (\ie, with a shared branch and a personalized branch), but uses a different strategy for federated model aggregation. 
Specifically, all parameters from the shared and personalized branches are sent to the central server and averaged. 
The server then sends back the averaged parameters to each site, serving as the initial parameters for the model on each site in the next communication round. 

(5) \textbf{FedProx}~\cite{li2020federated}. This method shares the same network architecture with SFGL and utilizes a different federated aggregation strategy. 
Specifically, the parameters received from each local site are averaged to obtain the global parameters. 
We then compute the $L2$ norm between the parameters of each site and the global parameters individually. 
These norms are then introduced as regularization terms into the loss function of each site (see Eq.~\eqref{EQ11}). 
This operation aims to mitigate the bias between global and local parameters during the optimization process, ultimately reducing parameter drift. 

\begin{figure*}[!t]
\setlength{\abovecaptionskip}{-4pt}
\setlength{\belowcaptionskip}{-2pt}
\setlength\abovedisplayskip{-1pt}
\setlength\belowdisplayskip{-1pt}
\centering
\includegraphics[width=1\textwidth]{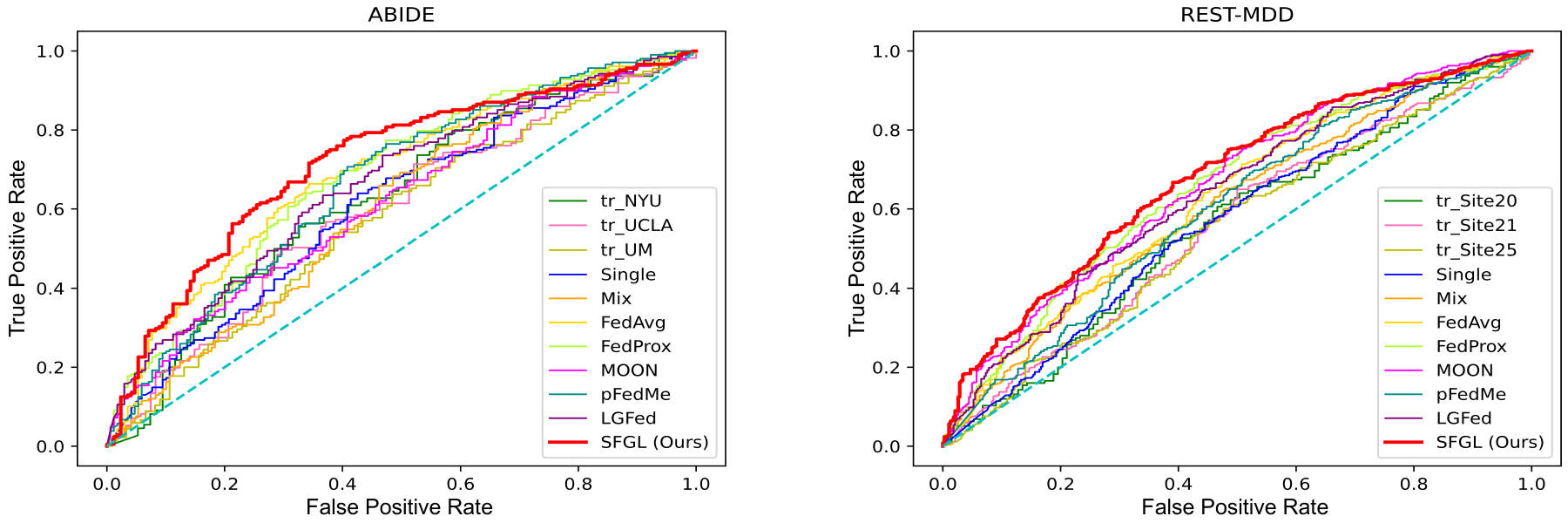}
\caption{The ROC curves plotted using different methods on two datasets. 
The left subplot is the ASD vs. NC classification performed on the ABIDE dataset, while the right subplot is the MDD vs. NC classification performed on the REST-MDD dataset.}
\label{ROC}
\end{figure*}

(6) \textbf{MOON}~\cite{li2021model}. Similar to FedAvg and FedProx, this method has the same network architecture as our SFGL but a different federated aggregation strategy. 
At each site, we consider the shared and personalized branches as a holistic model, and use the dynamic FCN sequence, vectorized demographic information and FCN as inputs to extract a 16-dimensional feature representation, called local representation, for disease prediction. 
The representation will be fed into a fully connected layer and a softmax layer for classification. 
Then, all parameters of the model at each site are sent to a server for average aggregation, and using the averaged parameters, a new 16-dimensional feature representation called global representation is generated at each site. 
We maximize the cosine similarity between the local and global representation while minimizing the cosine similarity between the local representation generated in the current communication round and the one generated in the previous round in order to leverage the similarity between model representations to correct the local training. 

(7) \textbf{pFedMe}~\cite{t2020personalized}. With the same network architecture as SFGL, this method treats the global model at the server side as a central point agreed upon by all sites and each local model as a point constructed by each site based on its heterogeneous data distribution. 
During parameter aggregation on the server side, a Moreau envelope is utilized as a regularization loss function to decouple the optimization process of the personalized model from the learning process of the global model, allowing the utilization of the global model to optimize the local model. 

(8) \textbf{LGFed}~\cite{liang2020think}. This method also shares the same network architecture with our SFGL. 
Different from SFGL, LGFed only sends the parameters of the last fully connected layer in the personalized branch to the central server for aggregation while other parameters remain at each local client. 
This enables joint learning of compact local representations on each site, and also reduces the number of communicated parameters for federated aggregation. 

\subsection{Experimental Results}
We report the experimental results of our SFGL and eight competing methods on ABIDE and REST-MDD in Tables~\ref{ABIDE_result}-\ref{MDD_result}. 
For ASD vs. NC classification on ABIDE, Table~\ref{ABIDE_result} shows the results of these methods on three sites (\ie, NYU, UCLA, UM), as well as the average results of these sites (called ``Average'') to evaluate the overall performance of FL or non-FL systems. 
Similarly, Table~\ref{MDD_result} shows classification results on three  sites (\ie, Site20, Site21, Site25) from REST-MDD.
For each method in \textbf{Cross} (\eg, tr\_NYU), the average result here is the average of two sites (\eg, UCLA and UM) that are used for test. 
In Fig.~\ref{ROC}, we plot the ROC curves of the average results of different methods on ABIDE and REST-MDD datasets. 
From Tables~\ref{ABIDE_result}-\ref{MDD_result} and Fig.~\ref{ROC}, we have the following interesting findings. 

\begin{itemize}
\item On one hand, for the tasks of ASD vs. NC classification on ABIDE and MDD vs. NC classification on REST-MDD, FL methods (\ie, FedAvg, FedProx, MOON, pFedMe, LGFed and SFGL) show significant improvements compared to the three non-FL methods (\ie, Cross, Single and Mix) in terms of average results. 
For example, as shown in Table~\ref{ABIDE_result} on ABIDE, our SFGL improves the average AUC by $12.3\%$ compared to the method training on a single site (\ie, Single), and achieves an improvement of $11.4\%$ compared to the method that directly mixes data from all sites for model training (\ie, Mix). 
On REST-MDD (see Table~\ref{MDD_result}), our method shows an average AUC improvement of $10\%$ and $7.2\%$ compared to the Single and Mix methods, respectively. 
These results demonstrate that FL enables multiple sites to collaboratively train models, which allows each site to leverage complementary knowledge from multiple sources, thus enhancing classification performance. 
    
\item On the other hand, our proposed SFGL generally outperforms the other five FL methods that do not consider site specificity (\ie, FedAvg, FedProx, MOON, pFedMe, and LGFed) in the two tasks. 
For example, on the ABIDE, the SFGL achieves an improvement of $2.7\%$ in terms of average ACC compared to FedProx and an improvement of $4.2\%$ compared to LGFed. 
On the REST-MDD, SFGL improves the average ACC results by $2.9\%$ and $2.1\%$ compared to FedAvg and MOON, respectively. 
The possible reason is that our SFGL that integrates a shared branch and a personalized branch can not only learn the spatiotemporally dynamic representation of fMRI, 
but also preserve the site-specific information from imaging and non-imaging views, thus achieving better performance in fMRI-based disorder diagnosis. 
\end{itemize}

\begin{table}[!tbp]
\setlength{\belowcaptionskip}{1pt}
\setlength{\abovecaptionskip}{1pt}
\setlength\abovedisplayskip{1pt}
\setlength\belowdisplayskip{1pt}
\renewcommand\arraystretch{1.1}
\centering
\caption{Results of statistical significance analysis by comparing the proposed SFGL and five federated learning methods.}
\scriptsize
\centering
\setlength{\tabcolsep}{0.6pt}
\label{t-test}
\begin{tabular*}{0.48\textwidth}{@{\extracolsep{\fill}} l|lcc }
\toprule
~~Dataset & \multicolumn{1}{c}{Pairwise Comparison} & \multicolumn{1}{c}{$p$-value} & $p< 0.05$ \\ 
\hline
\multirow{5}{*}{~~ABIDE}  & SFGL vs. FedAvg  & $1.042\times 10^{-2}$         & $\ast$    \\
                          & SFGL vs. FedProx & $7.734\times 10^{-3}$         & $\ast$    \\
                          & SFGL vs. MOON    & $9.976\times 10^{-3}$         & $\ast$    \\
                          & SFGL vs. pFedMe  & $9.578\times 10^{-2}$         &      \\
                          & SFGL vs. LGFed   & $1.475\times 10^{-2}$         & $\ast$    \\ 
\hline
\multirow{5}{*}{~~REST-MDD} & SFGL vs. FedAvg  & $1.382\times 10^{-2}$       & $\ast$    \\
                            & SFGL vs. FedProx & $3.906\times 10^{-7}$       & $\ast$    \\
                            & SFGL vs. MOON    & $3.960\times 10^{-11}$      & $\ast$    \\
                            & SFGL vs. pFedMe  & $1.157\times 10^{-7}$       & $\ast$    \\
                            & SFGL vs. LGFed   & $1.695\times 10^{-5}$       & $\ast$    \\
\bottomrule
\end{tabular*}
\end{table}

\begin{figure*}[!t]
\setlength{\abovecaptionskip}{-1pt}
\setlength{\belowcaptionskip}{-2pt}
\setlength\abovedisplayskip{-1pt}
\setlength\belowdisplayskip{-1pt}
\centering
\includegraphics[width=0.98\textwidth]{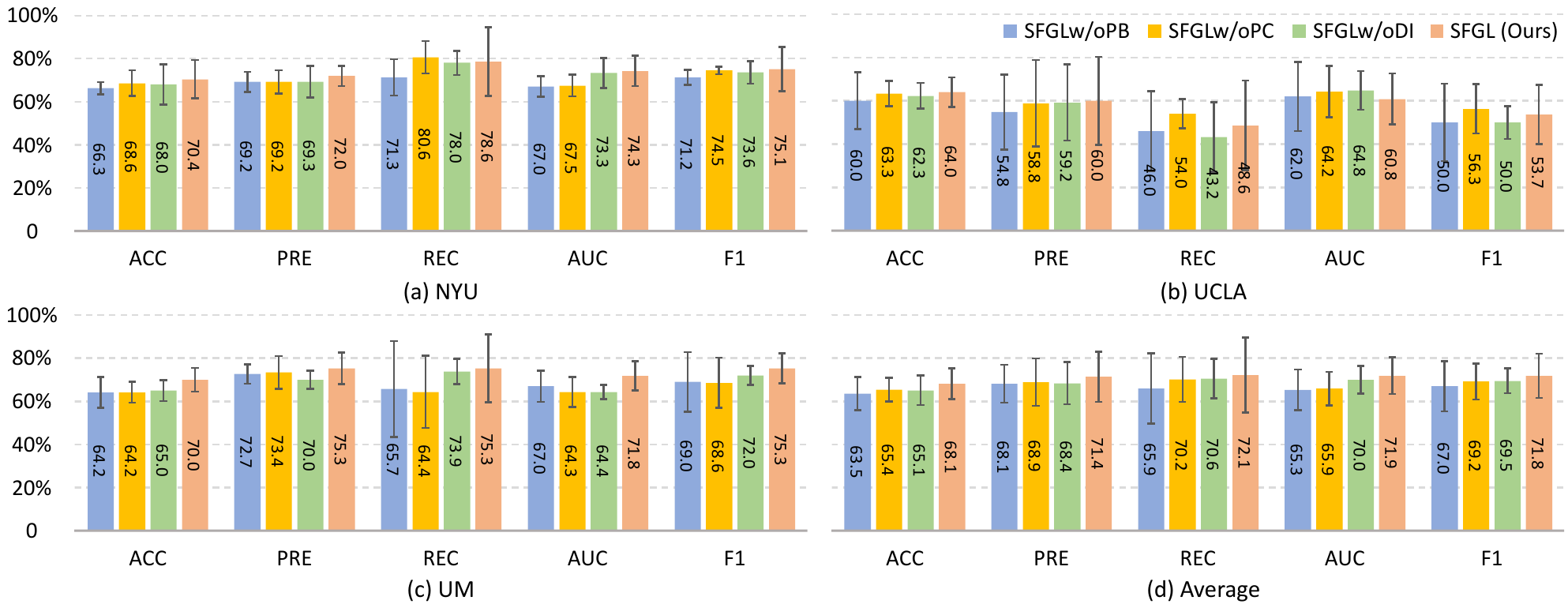}
\caption{Performance of our SFGL and its three variants in the task of ASD vs. NC classification on ABIDE.}
\label{wo_ABIDE}
\end{figure*}

\begin{table*}[!tbp]
\setlength{\belowcaptionskip}{1pt}
\setlength{\abovecaptionskip}{1pt}
\setlength\abovedisplayskip{1pt}
\setlength\belowdisplayskip{1pt}
\scriptsize
\tiny
\centering
\renewcommand{\arraystretch}{1.1}
\setlength{\tabcolsep}{2pt}
\caption{Results (\%) of SFGL and its three variants in MDD vs. NC classification on the REST-MDD in terms of mean$\pm$standard deviation. Best results are shown in bold.}
\begin{tabular*}{1\textwidth}{@{\extracolsep{\fill}} l|ccccc|ccccc|ccccc|ccccc}
\toprule

\multirow{2}{*}{~Method}
& \multicolumn{5}{c|}{Site20} & \multicolumn{5}{c|}{Site21} & \multicolumn{5}{c|}{Site25} 
& \multicolumn{5}{c}{Average} \\                                                   
\cline{2-21} 
&ACC & PRE & REC & AUC & F1 & ACC & PRE & REC & AUC & F1 & ACC & PRE & REC & AUC & F1 & ACC & PRE & REC & AUC & F1 \\ 
\hline
\multirow{1}{*}{~SFGLw/oPB}
& \shortstack[c]{60.6$\pm$5.2} & \shortstack[c]{59.6$\pm$3.4} & \shortstack[c]{71.7$\pm$10.8} & \shortstack[c]{63.1$\pm$4.3} & \shortstack[c]{67.0$\pm$5.8} & \shortstack[c]{57.1$\pm$6.7} & \shortstack[c]{60.0$\pm$5.3} & \shortstack[c]{63.9$\pm$10.4} & \shortstack[c]{61.0$\pm$12.6} & \shortstack[c]{62.1$\pm$6.1} & \shortstack[c]{53.3$\pm$5.8} & \shortstack[c]{60.7$\pm$5.1} & \shortstack[c]{57.3$\pm$10.6} & \shortstack[c]{56.9$\pm$8.6} & \shortstack[c]{59.0$\pm$7.4} & \shortstack[c]{57.0$\pm$5.9} & \shortstack[c]{60.0$\pm$4.6} & \shortstack[c]{71.7$\pm$10.5} & \shortstack[c]{61.1$\pm$8.5} & \shortstack[c]{56.3$\pm$6.4}\\

\multirow{1}{*}{~SFGLw/oPC}
& \shortstack[c]{60.0$\pm$5.4} & \shortstack[c]{61.0$\pm$4.6} & \shortstack[c]{66.7$\pm$7.6} & \shortstack[c]{63.2$\pm$4.4} & \shortstack[c]{63.7$\pm$5.9} & \shortstack[c]{59.6$\pm$6.4} & \shortstack[c]{63.2$\pm$5.0} & \shortstack[c]{64.0$\pm$17.4} & \textbf{\shortstack[c]{63.0$\pm$7.3}} & \shortstack[c]{63.6$\pm$10.3} & \shortstack[c]{61.1$\pm$6.6} & \textbf{\shortstack[c]{70.2$\pm$4.2}} & \shortstack[c]{58.4$\pm$11.1} & \shortstack[c]{62.6$\pm$3.7} & \shortstack[c]{63.8$\pm$8.1} & \shortstack[c]{60.2$\pm$6.2} & \shortstack[c]{62.9$\pm$4.6} & \shortstack[c]{64.6$\pm$12.0} & \shortstack[c]{62.6$\pm$5.1} & \shortstack[c]{63.7$\pm$8.1}\\

\multirow{1}{*}{~SFGLw/oDI}
& \shortstack[c]{61.3$\pm$2.4} & \shortstack[c]{63.1$\pm$4.0} & \shortstack[c]{64.9$\pm$11.3} & \shortstack[c]{66.2$\pm$3.7} & \shortstack[c]{63.9$\pm$4.2} & \shortstack[c]{58.3$\pm$3.8} & \shortstack[c]{61.7$\pm$4.0} & \shortstack[c]{63.9$\pm$20.7} & \shortstack[c]{62.9$\pm$6.7} & \shortstack[c]{62.8$\pm$10.8} & \shortstack[c]{61.8$\pm$5.2} & \shortstack[c]{64.3$\pm$4.4} & \shortstack[c]{74.1$\pm$7.4} & \shortstack[c]{62.7$\pm$6.2} & \shortstack[c]{69.4$\pm$5.5} & \shortstack[c]{60.5$\pm$3.8} & \shortstack[c]{63.3$\pm$4.1} & \shortstack[c]{66.5$\pm$13.2} & \shortstack[c]{65.3$\pm$5.5} & \shortstack[c]{64.9$\pm$6.8}\\
\hline
\multirow{1}{*}{~SFGL (Ours)}
& \textbf{\shortstack[c]{64.4$\pm$2.4}} & \textbf{\shortstack[c]{64.5$\pm$2.9}} & \textbf{\shortstack[c]{72.7$\pm$10.2}} 
& \textbf{\shortstack[c]{68.9$\pm$4.1}} & \textbf{\shortstack[c]{68.3$\pm$4.1}} & \textbf{\shortstack[c]{62.0$\pm$4.3}} 
& \textbf{\shortstack[c]{63.7$\pm$4.2}} & \textbf{\shortstack[c]{65.1$\pm$7.6}} & \shortstack[c]{62.3$\pm$6.9} 
& \textbf{\shortstack[c]{64.4$\pm$4.1}} & \textbf{\shortstack[c]{62.5$\pm$6.0}} & \shortstack[c]{64.5$\pm$3.4} 
& \textbf{\shortstack[c]{79.8$\pm$11.4}} & \textbf{\shortstack[c]{69.2$\pm$7.9}} & \textbf{\shortstack[c]{71.4$\pm$5.6}}
& \textbf{\shortstack[c]{62.9$\pm$4.7}} & \textbf{\shortstack[c]{64.3$\pm$3.4}} & \textbf{\shortstack[c]{72.6$\pm$9.7}}
& \textbf{\shortstack[c]{67.8$\pm$6.3}} & \textbf{\shortstack[c]{68.2$\pm$4.6}}\\ 

\bottomrule
\end{tabular*}
\label{wo_MDD}
\end{table*}

\subsection{Statistical Significance Analysis}
Through paired-sample $t$-test, we further calculate the differences of predicted probability distribution between our SFGL and five competing FL methods. 
The test statistic for the paired-sample $t$-test can be represented as $t = \frac{\overline{x}_{\mathrm{diff}}}{{s_{\mathrm{diff}}}/{\sqrt{n}}}$, where $\overline{x}_{\mathrm{diff}}$ represents the mean of the sample differences, $s_{\mathrm{diff}}$ represents the standard deviation of the sample differences, and $n$ is the number of pairs of samples. 
The corresponding $p$-values for the $t$-test are shown in Table~\ref{t-test}. 
We set the significance level threshold as $0.05$, where if the $p$-value is less than $0.05$, we consider our method shows a significant difference compared to the competing method, represented as ``$\ast$'' in Table~\ref{t-test}. 
From Table~\ref{t-test}, we can find that the SFGL  is statistically significantly different from the five FL methods in most cases, which further validates the superiority of SFGL in fMRI-based disorder classification. 

\section{Discussion}\label{S6}
\subsection{Ablation Study}\label{Ablation}
To validate the effectiveness of several key components in our SFGL, we compare the SFGL with its three degenerated variants called \textbf{SFGLw/oPB}, \textbf{SFGLw/oPC}, and \textbf{SFGLw/oDI}, respectively. 
Specifically, the SFGLw/oPB only uses the shared branch and discards the personalized branch, where all parameters of local models are sent to the server for federal aggregation. 
In SFGLw/oPC, the personalized branch of each local model only utilizes demographic information. 
Similarly, SFGLw/oDI only considers the imaging information (\ie, FCN) in the personalized branch. 
Although the personalized branches in SFGLw/oPC and SFGLw/oDI are degenerated, their parameters still do not participate in federated aggregation at the server side and remain updated at each local client. 
In Fig.~\ref{wo_ABIDE} and Table~\ref{wo_MDD}, we report the experimental results of the four methods in two classification tasks on ABIDE and REST-MDD. 

It can be observed from Fig.~\ref{wo_ABIDE} and Table~\ref{wo_MDD} that, in most cases, SFGL outperforms its three variants that discard partial or complete information of the personalized branch (\ie, SFGLw/oPB, SFGLw/oPC, and SFGLw/oDI) in both ASD vs. NC and MDD vs. NC classification tasks. 
This validates the effectiveness of the designed personalized branch in preserving site-specificity information. 
Besides, two variants with partially personalized branches retained (\ie, SFGLw/oPC and SFGLw/oDI) is generally inferior to our SFGL which employs the complete personalized branch. 
The results indicate that FCN and demographic information in the personalized branch can provide complementary site-specific information from imaging and non-imaging views to help boost classification performance. 

\subsection{Influence of Balancing Coefficient}
\begin{figure}[!t]
\setlength{\abovecaptionskip}{-1pt}
\setlength{\belowcaptionskip}{-2pt}
\setlength\abovedisplayskip{-1pt}
\setlength\belowdisplayskip{-1pt}
\centering
\includegraphics[width=0.49\textwidth]{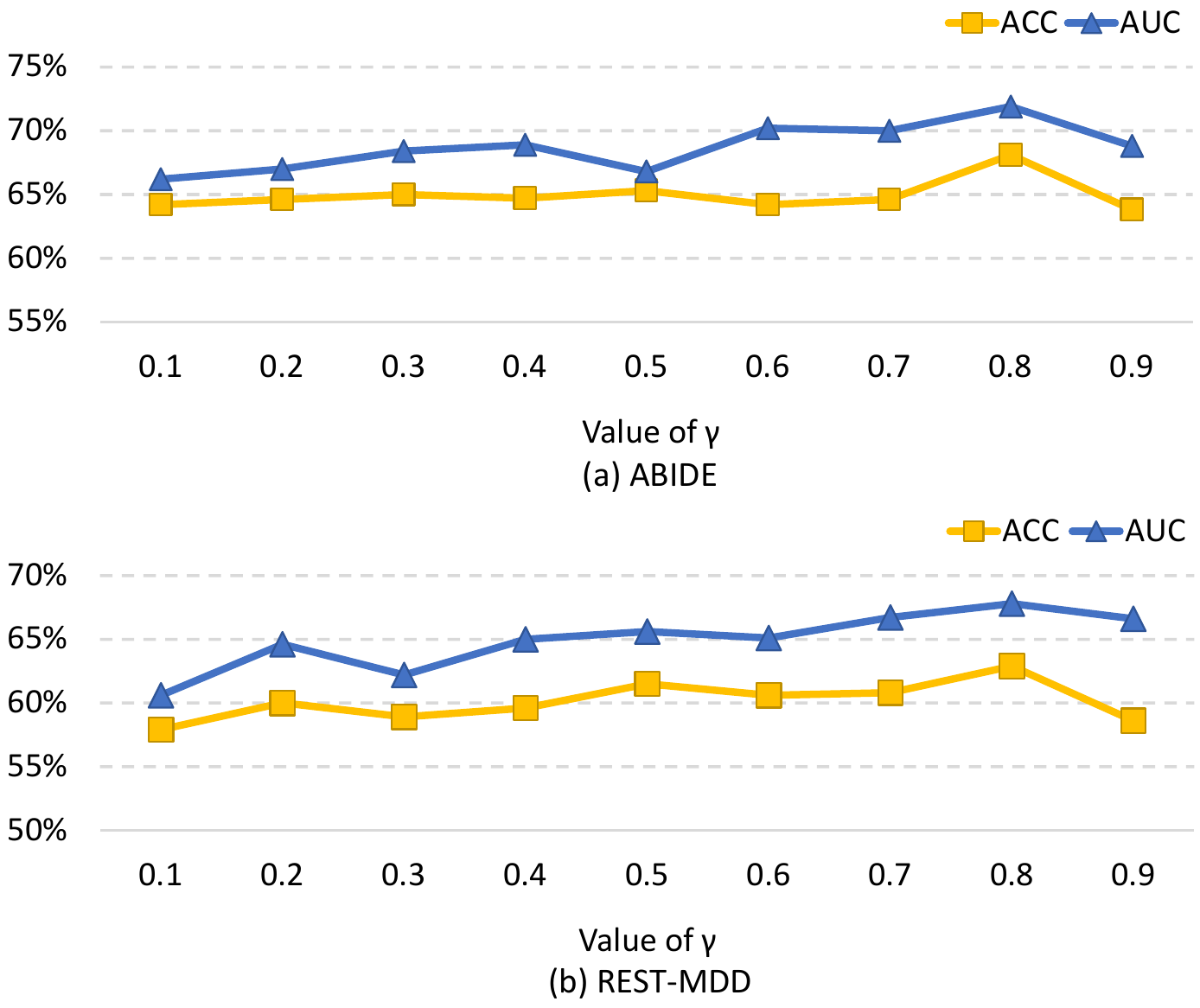}
\caption{Results of our SFGL with different values of $\gamma$ in Eq.~\eqref{EQ6} in the task of (a) ASD vs. NC classification on ABIDE and (b) MDD vs. NC classification on REST-MDD. }
\label{gamma}
\end{figure}

In Eq.~\eqref{EQ6}, the balancing coefficient $\gamma $ adjusts the contribution of features from the shared branch and the personalized branch to the final prediction. 
A larger value of $\gamma$ means a larger contribution of the personalized branch, while a smaller value indicates a larger contribution of the shared branch. 
In particular, when $\gamma =0$, it means that only the shared branch is used, which is equivalent to the SFGLw/oPB variant in Section~\ref{Ablation}. 
In the main experiment, we empirically set $\gamma=0.8$. 
To investigate the impact of different balancing coefficients, we vary the value of $\gamma $ within the range of $\{0.1, 0.2, \cdots, 0.9\}$ and report the results of SFGL in the two tasks on ABIDE and REST-MDD in Fig.~\ref{gamma}. 

From Fig.~\ref{gamma}, it can be found that when the value of $\gamma $ is very large (\eg, $\gamma=0.9$), the SFGL only achieves the average ACC of $58.6\%$ for MDD vs. NC classification on REST-MDD. 
The possible reason is that the SFGL can not effectively leverage collective knowledge from multiple sites when the SFGL excessively focuses on the personalized branch, thus yielding suboptimal classification performance. 
On the contrary, if $\gamma $ is set to be a very small value (\eg, $\gamma=0.4$), the SFGL still performs poorly with AUC=$68.9\%$ on ABIDE and AUC=$65.0\%$ on REST-MDD.
This may be due to inadequate preservation of site-specific information, resulting in poor performance. 

\subsection{Influence of Local Training Epoch}
\begin{figure}[!t]
\setlength{\abovecaptionskip}{-1pt}
\setlength{\belowcaptionskip}{-2pt}
\setlength\abovedisplayskip{-1pt}
\setlength\belowdisplayskip{-1pt}
\centering
\includegraphics[width=0.49\textwidth]{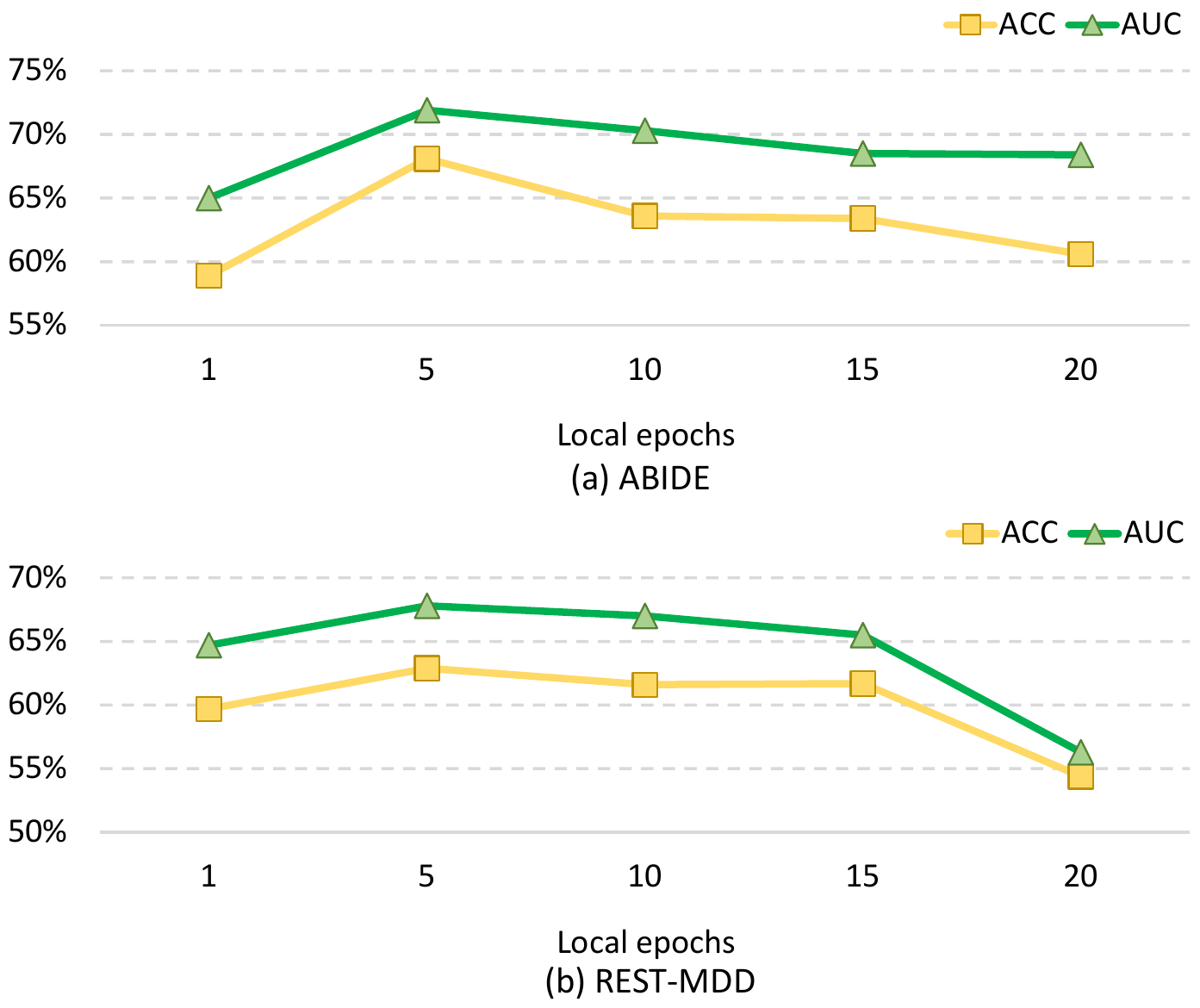}
\caption{Results of SFGL with different local epochs in the task of (a) ASD vs. NC classification on ABIDE and (b) MDD vs. NC classification on REST-MDD. }
\label{LE}
\end{figure}

In this work, the local epoch $E$ denotes the number of training iterations on each client before the updated model parameters are sent back to the central server for aggregation, which is also a crucial hyperparameter in the proposed SFGL. 
In the main experiments, we empirically set the local epoch $E$ as $5$. 
To explore the impact of different local epochs, we vary the value of $E$ within the range of $\left \{ 1, 5, 10, 15, 20 \right \} $, and record the results of SFGL in the two tasks on ABIDE and REST-MDD in Fig.~\ref{LE}. 
From Fig.~\ref{LE}, we can see that the SFGL achieves worse performance when the local epoch is very small (\eg, $E=1$) under the same communication rounds. 
This is likely because frequent updates from local clients might hinder model stabilization when a small local epoch is used, thus degrading model performance. 
And the SFGL consistently achieves the best performance on ABIDE and REST-MDD with $E=5$. 
Besides, it can be observed that the performance of SFGL is inferior when a large local epoch (\ie, 20) is used. 
This may be due to that using a large local epoch could lead to the parameter drift issue (\ie, misalignment between local and global optima). 

\subsection{Influence of Different Backbones in Shared Branch}
In the shared branch of the SFGL, we use STAGIN which integrates GIN and Transformer as the backbone to extract dynamic graph features. 
To study how different backbones affect the performance of our SFGL, we replace the STAGIN with several popular GNNs, including (1) graph attention network (GAT)~\cite{velickovic2017graph}, (2) graph convolutional network (GCN)~\cite{kipf2016semi}, (3) graph isomorphism network (GIN)~\cite{xu2018powerful} and (4) spatio-temporal graph convolutional network (STGCN)~\cite{gadgil2020spatio}. 
Specifically, the GAT, GCN, and GIN use static graphs as input, and their static graph construction strategy is the same as each of the dynamic graphs in SFGL (see Section~\ref{graph}). 
But in STGCN, we use BOLD signal time series as the feature input, and the overall PC matrix calculated from concatenating the BOLD signals of all subjects is utilized as the adjacency matrix. 
In each STGC layer, the output channel is 64, and the number and size of spatial convolutional kernels are 1 and 64. 
For a fair comparison, all the backbones use two GNN layers, which is consistent with the STAGIN used in SFGL. 
The experimental results of our SFGL with five different backbones in two prediction tasks on ABIDE and REST-MDD are reported in Table~\ref{backbone}. 

\begin{table*}[!tbp]
\setlength{\belowcaptionskip}{1pt}
\setlength{\abovecaptionskip}{1pt}
\setlength\abovedisplayskip{1pt}
\setlength\belowdisplayskip{1pt}
\centering
\renewcommand{\arraystretch}{1}
\caption{Results (\%) of SFGL using different backbones in the shared branch in the task of ASD vs. NC classification on ABIDE and MDD vs. NC classification on REST-MDD in terms of mean$\pm$standard deviation. The best results are shown in bold.}
	\scriptsize
	\centering
\setlength{\tabcolsep}{2pt}	
    \begin{tabular*}{1\textwidth}{@{\extracolsep{\fill}} l |ccccc|ccccc}
			\toprule
		    \multirow{2}{*}{~Backbone}      &\multicolumn{5}{c|}{ASD vs. NC classification on ABIDE}    &\multicolumn{5}{c}{MDD vs. NC classification on REST-MDD}     \\
		    \cline{2-11}
		    & ACC     & PRE    & REC     & AUC     &F1     & ACC     & PRE    & REC     & AUC     &F1    \\
			\hline
           ~GAT~   &63.9$\pm$5.0      &64.8$\pm$7.1      &76.4$\pm$16.0      &68.2$\pm$7.5      &70.2$\pm$7.7
                      &61.7$\pm$2.8     &64.6$\pm$1.9      &71.3$\pm$6.0      &66.7$\pm$5.6      &67.8$\pm$3.2\\
           ~GCN~    &65.0$\pm$9.9     &66.8$\pm$16.7      &\textbf{78.3$\pm$19.5}      &69.3$\pm$10.0      &71.2$\pm$12.8
                      &61.1$\pm$8.4     &\textbf{65.1$\pm$7.7}      &68.0$\pm$14.5      &65.6$\pm$8.4      &66.6$\pm$10.6\\ 
           ~GIN~    &63.8$\pm$8.0     &65.2$\pm$6.7      &76.4$\pm$16.4      &65.7$\pm$7.0      &70.5$\pm$9.9
                      &61.5$\pm$6.8     &62.6$\pm$5.7      &71.5$\pm$11.6      &66.3$\pm$8.9      &66.8$\pm$7.5\\    
           ~STGCN~    &65.2$\pm$6.1     &65.6$\pm$5.4      &77.8$\pm$13.6      &67.3$\pm$6.5      &71.2$\pm$8.1
                      &61.7$\pm$2.4     &64.8$\pm$4.0      &69.3$\pm$7.3      &67.7$\pm$5.8      &67.0$\pm$3.2\\
           ~STAGIN~    &\textbf{68.1$\pm$7.1}     &\textbf{71.4$\pm$11.7}      &72.1$\pm$17.4      &\textbf{71.9$\pm$8.5}      &\textbf{71.8$\pm$10.3}
                      &\textbf{62.9$\pm$4.7}     &64.3$\pm$3.4      &\textbf{72.6$\pm$9.7}      &\textbf{67.8$\pm$6.3}      &\textbf{68.2$\pm$4.6}\\
        \bottomrule
		\end{tabular*}
	\label{backbone}
\end{table*}

\begin{figure*}[!t]
\setlength{\abovecaptionskip}{-1pt}
\setlength{\belowcaptionskip}{-1pt}
\setlength\abovedisplayskip{-1pt}
\setlength\belowdisplayskip{-1pt}
\centering
\includegraphics[width=1\textwidth]{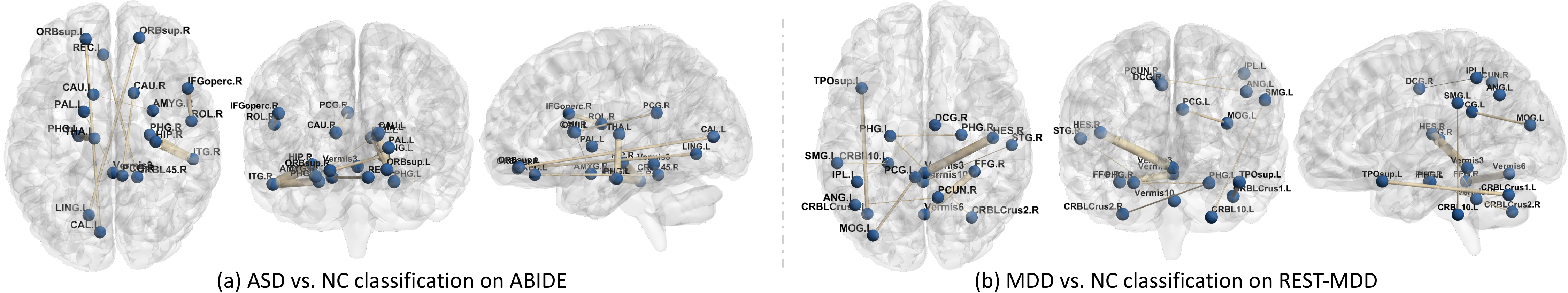}
\caption{Visualization of the top 10 most discriminative functional connections detected by the proposed SFGL in tasks of (a) ASD vs. NC classification on NYU from ABIDE~\cite{di2014autism} and (b) MDD vs. NC classification on Site 20 from REST-MDD~\cite{yan2019reduced}. }
\label{bio}
\end{figure*}

From Table~\ref{backbone}, we can observe that SFGL with two spatio-temporal GNN backbones (\ie, STGCN and STAGIN) generally outperform the other three static backbones that do not consider temporal dynamics in fMRI series (\ie, GAT, GCN and GIN) on ABIDE and REST-MDD datasets. 
This indicates that using spatio-temporal GNN backbones helps extract more discriminative features via modeling both the spatial and temporal patterns of fMRI data, thus enhancing model performance. 
Furthermore, SFGL with the STAGIN backbone achieves better results compared with STGCN which uses 1-D convolution for temporal feature abstraction in most cases. 
For instance, STAGIN exhibits superiority by $2.9\%$ and $1.2\%$ in terms of the ACC metric on ABIDE and REST-MDD, respectively. 
The possible reason is that the STAGIN can capture long-range temporal dependencies of fMRI series via transformer encoder to further facilitate fMRI-based disease diagnosis. 

\subsection{Interpretable Biomarker Analysis}
Identifying interpretable biomarkers (\eg, functional connections and ROIs) associated with brain diseases can provide insights into underlying neural mechanisms and help early detection and targeted treatment of brain disorders. 
To enhance the interpretability of our proposed SFGL, we utilize a guided back-propagation gradient-based approach~\cite{springenberg2014striving,selvaraju2017grad} to detect the discriminative functional connections and ROIs on ABIDE and MDD datasets. 
The guided back-propagation gradient is formulated as: 
\begin{equation}
g_k^{c}=\mathrm{ReLU} \left ( \frac{\partial y^c }{\partial x_k}  \right ), 
\label{EQ16}
\end{equation}
where $c \in \left \{ 0,1, \cdots, C-1 \right \} $ denotes a specific category, $C$ denotes the number of categories, $y^c$ is the logit score for class $c$ before the softmax layer, and $x_k$ is the $k$-th feature of the input. 

Specifically, taking the ABIDE as an example, we select all ASD patients from NYU for biomarker analysis. 
For a given subject, in the proposed shared branch of our SFGL, we first use Eq.~\eqref{EQ16} to compute the guided gradient matrix based on predicted logit score and input FCN $\left \{ X(t) \right \}_{t=1}^T $ for each time segment, and then average the gradient matrices of all segments to yield shared gradient matrix $G_s$. 
In the personalized branch, we use Eq.~\eqref{EQ16} to obtain a personalized gradient matrix, with a predicted logit score and the full-time FCN as input. 
The guided gradient matrix $G$ for each subject is obtained via the weighted sum of $G_s$ and $G_p$, formulated as $G= \gamma G_p+(1-\gamma )G_s$, where the weight $\gamma$ is same as that in Eq.~\eqref{EQ6}. 
Finally, we average the $G$ matrices of all ASD patients to select the top $10$ brain functional connectivity corresponding to the largest gradients. 
Similarly, we select all MDD patients from Site 20 of the REST-MDD dataset for biomarker analysis. 
In Fig.~\ref{bio}, we map the identified FCs and ROIs to the brain space and visualize them using BrainNet Viewer~\cite{xia2013brainnet}. 

As illustrated in Fig.~\ref{bio}~(a), the proposed SFGL identifies several discriminative brain ROIs, such as the right hippocampus (HIP.R), right amygdala (AMYG.R), left (right) parahippocampal gyrus (PHG.L, PHG.R), left (right) caudate nucleus (CAU.L, CAU.R), left thalamus (THA.L), left pallidum (PAL.L), which have been mentioned in previous studies on ASD patients~\cite{baron2000amygdala, padmanabhan2017default, roy2021anterior}. 
In Fig.~\ref{bio}~(b), the SFGL also identifies several brain ROIs associated with MDD identification, such as left (right) parahippocampal gyrus (PHG.L, PHG.R), right fusiform gyrus (FFG.R), right superior temporal gyrus (STG.R), left posterior cingulate gyrus (PCG.L), right precuneus (PCUN.R), which is consistent with previous MDD research~\cite{hermesdorf2016major, wang2015interhemispheric, guo2013decreased}. 
These results further demonstrate the  reliability of our SFGL in discovering disease-related biomarkers. 

\subsection{Limitations and Future Work}
This work has some limitations that need to be taken into consideration. 
\emph{First}, in the shared branch of SFGL, we only capture the dependencies between brain ROIs during graph spatial feature learning, without considering the potential relationships between subjects. 
In future work, we will incorporate pairwise or triplet relationships among subjects~\cite{li2022functional,yao2021mutual} into the proposed framework to further enhance discriminative ability of learned features. 
\emph{Second}, all sites employ the same network architecture of the personalized branch in SFGL. 
Considering the parameters of the personalized branch only remain updated at each local client, different sites may benefit more from using different personalized branches based on their unique data characteristics, which will be a future research direction. 
\emph{In addition}, in the current work, we linearly fuse features output by the shared and the personalized branches (see Eq.~\eqref{EQ6}). 
It is interesting to design more advanced feature fusion strategies (\eg, via Transformer~\cite{han2022survey}) to integrate the features of different branches, and this will be one of our future works. 

\section{Conclusion}\label{S7}
In this paper, we propose a specificity-aware federated graph learning (SFGL) framework for multi-site fMRI analysis and brain disorder diagnosis. 
The SFGL consists of a shared branch to facilitate cross-site knowledge sharing, and a personalized branch to preserve site specificity information. 
Specifically, we employ GIN and transformer layers to capture the spatio-temporal information of brain FCNs in the shared branch. 
In the designed personalized branch, we extract site-specific features from both imaging (\ie, static FCN) and non-imaging (\ie, demographic information) views. 
Besides, a novel federal aggregation strategy is developed in the proposed SFGL, where the parameters of the shared branch are sent to a server while the parameters of personalized branch remain updated at each local client/site. 
Extensive experiments validate the superiority of the SFGL in automated brain disorder identification and biomarker detection based on multi-site fMRI data. 

\section*{Declaration of Competing Interest}
The authors declare that they have no known competing financial interests or personal relationships that could have appeared to influence the work reported in this paper. 

\section*{CRediT Authorship Contribution Statement}
{\textbf{J.~Zhang}}: Methodology, Software, Writing - original draft. 
{\textbf{Q.~Wang}}: Data collection, Writing - review \& editing. 
{\textbf{X.~Wang}}: Data collection, Writing - review \& editing. 
{\textbf{L.~Qiao}}: Writing - review \& editing,  Supervision. 
{\textbf{M.~Liu}}: Conceptualization, Validation, Writing - review \& editing, Supervision. 

\bibliographystyle{model2-names.bst}
\biboptions{authoryear}
\bibliography{main}

\end{document}